\documentclass[a4paper,fleqn]{cas-sc}

\usepackage[authoryear,longnamesfirst]{natbib}

\def\tsc#1{\csdef{#1}{\textsc{\lowercase{#1}}\xspace}}
\tsc{WGM}
\tsc{QE}
\tsc{EP}
\tsc{PMS}
\tsc{BEC}
\tsc{DE}
\newcommand{\Lop}{\ensuremath{{\cal L}}}  
\newcommand{\pS}{\ensuremath{{\cal M}}}   

\begin{document}
\let\WriteBookmarks\relax
\def\floatpagepagefraction{1}
\def\textpagefraction{.001}
\shorttitle{Chaotic fields out of equilibrium are observable independent}
\shortauthors{D. Lippolis}

\title [mode = title]{Chaotic fields out of equilibrium are observable independent}                      



\author{D. Lippolis}[
                        orcid=0000-0002-2817-0859]
\ead{domenico@ujs.edu.cn}
\ead[url]{https://cns.gatech.edu/~domenico/}


\affiliation{organization={School of Mathematical Sciences, Jiangsu University},
                city={Zhenjiang},
                postcode={212013}, 
                state={Jiangsu},
                country={China}}









\begin{abstract}
Chaotic dynamics is always characterized by swarms of unstable trajectories, unpredictable individually, and thus generally studied statistically. It is often the case that such phase-space densities relax exponentially fast to a limiting distribution, that rules the long-time average of every observable of interest. Before that asymptotic time scale, the statistics of chaos is generally believed to depend on both the initial conditions and the chosen observable. I show that this is not the case for a widely applicable class of models, that feature a phase-space (`field') distribution common to all pushed-forward or integrated observables, while the system is still relaxing towards statistical equilibrium or a stationary state. This universal profile is determined by both leading and first subleading eigenfunctions of the transport operator (Koopman or Perron-Frobenius) that maps phase-space densities forward or backward in time.

\end{abstract}


\begin{highlights}
\item Distinct time-dependent observables advected in the phase space by a chaotic dynamical system and before relaxation to statistical equilibrium exhibit common features
\item The phase-space profiles of the dynamical observables at intermediate timescales are determined by the first two eigenfunctions of the Perron-Frobenius/Koopman operator 
\item Theoretical predictions are verified numerically on expanding (Bernoulli map), hyperbolic (perturbed cat map), and non-hyperbolic (Hamiltonian H\'enon map) paradigms of chaos
\end{highlights}


\maketitle

\section{Introduction}

As a fact of nature, chaos is unpredictable, due to the extreme sensitivity of trajectories to 
initial conditions, and the finite precision of our computers. Impossible as it is to solve the problem of motion, practitioners resort to the tools of statistical mechanics to estimate averages and fluctuations of observables of interest (energy, diffusion constant, Lyapunov exponents, vorticity, etc.).  That feat is achievable 
for long-time evolution, provided that the system in exam be ergodic, and that it asymptotically 
relax to a limiting distribution usually called natural measure, the weight function of phase-space averages. 
  
The natural measure (or invariant density) is the leading eigenfunction
of the Perron-Frobenius transport operator $\Lop_t$~\cite{BratJorg}, that evolves  densities in time throughout the phase space, and whose leading eigenvalue 
determines the escape rate.
As the resolvent to the Liouville equation,  this operator is linear, but infinite dimensional~\cite{dasBuch}.
Especially in the Koopman form (adjoint of Perron-Frobenius), transport operators have found countless applications in recent years~\cite{Koopmanism,Deco,Raissa,Moreno},  
thanks to the genericity of the Markov property throughout modelling~\cite{Krauth}, and so have their  
finite-dimensional discretizations~\cite{Klus,Lucar,Andre}, which lend themselves to a faster time-iteration and 
a practical spectral analysis.

Spectrum wise, the first subleading eigenvalue yields information on the speed of relaxation of the system to
statistical equilibrium or steady state, and on the typical time scale of the decay of correlations. In one-dimensional chaos,
the second eigenvalue of the Perron-Frobenius operator has been related to the Lyapunov exponent of the dynamics~\cite{Slipan13}.
The role of the corresponding eigenfunction was first investigated in the context of passive fluid flows, where the term `strange eigenmode' was coined~\cite{Pierrehumb}.
It was surmised~\cite{PikPop} and later proven~\cite{LiuHaller} that the density of tracers in the fluid exhibits phase-space profiles which reflect
the second eigenfunction of the advection/diffusion differential operator, before the flow relaxes to equilibrium.   
Similar analogies between the density of phase-space trajectories and the non-trivial spectrum of the Perron-Frobenius operator 
were made in the context of discretization schemes~\cite{DellJunge} for the latter. In particular, the second eigenfunction of the 
transfer operator was related to quasi-invariant sets and almost cyclic behavior~\cite{Froy07}. Recently,  
the second eigenfunction of the Perron-Frobenius operator was shown to exhibit similar patterns to the distribution of finite-time Lyapunov exponents, that
is the field of local instabilities, in strong chaos~\cite{Kensuke,ClassScars} and in two-dimensional vector fields~\cite{RossBollt}.

The present report extends these findings by demonstrating that the first two eigenfunctions of the transfer operator spectrum govern not only the distribution of the Lyapunov exponents, but also that of  
\textit{any} field derived from numerically integrated trajectories (`Lagrangian' point of view in fluid dynamics), that is the phase-space distribution\footnote{In what follows, the word distribution is understood as phase-space function, say $a(x)$, and not as probability density function.} 
of an observable mapped forward (or backward)
in time, before relaxation to equilibrium or a stationary state.  

The formal analysis revolves around time-dependent dynamical averages, initially defined to
compute expectation values, but successively transformed to full distributions in the phase space,
more informative than single statistical moments that depend on initial conditions.  
 Next, the time evolution of the phase-space densities that serve as weight functions for the dynamical averages are expanded in terms of the eigenfunctions of the transfer operator. That  
 connects the obervables to the leading and subleading eigenfunctions of the spectrum,
 the central result of this work. The observable-independent behavior of the field profiles stems from their connection to these eigenfunctions, and it also extends to the so-called 
 time-integrated observables, such as finite-time Lyapunov exponents, or diffusion constants.
 Moreover, the derived formalism and results apply to chaotic systems with weak noise, besides purely 
 deterministic ones. The theory is validated analytically on the Bernoulli map, and
 numerically on
 the perturbed cat map, and the Hamiltonian H\'enon map.

\section{Time-dependent statistics of chaotic observables}
\subsection{Mapping phase-space averages}

Given an observable $a(x)$ defined as a function or an operator in a bounded region $\pS$ of the phase space, its expectation value may be written as
\begin{equation}
\langle a \rangle = \frac{1}{\mu_\pS(\rho)}\int_\pS a(x)\rho(x)\,dx\,,
\label{expa}
\end{equation}
where $\rho(x)$ is the phase space density, that is the frequency with which trajectories visit $\pS$, while $\mu_\pS(\cdot)$ is the Lebesgue measure over $\pS$. The observable $a$ is assumed smooth~\footnote{How smooth is discussed in Appendix~\ref{AppTrunc}.},  and like the density $\rho$, it is  a function of the dynamical variable $x$, that is advected by a
flow $f^t(x)$ as time proceeds. As a result, the expectation value~(\ref{expa}) itself depends on time as
 \begin{equation}
\langle a^t \rangle =  \frac{1}{\mu_\pS(\rho^t)}\int_\pS a(f^t(x))\rho(x)\,dx\,.
\label{expf}
\end{equation}
The pushed-forward (`downstream') observable $a(f^t(x))$ may also be expressed through the action of the Koopman operator,
$\left[\Lop^\dagger_t a\right](x) = a(f^t(x))$, while $\mu_\pS(\rho^t)=\int_\pS dx\left[\Lop_t \rho\right](x)$, the fraction of trajectories that have not escaped $\pS$ by time $t$. The Koopman operator is \textit{linear} and rewrites the integral in  equation~(\ref{expf}) as
\begin{equation}
\int_\pS \left[\Lop^\dagger_t a\right](x)\rho(x)\,dx\, =
\int_\pS a(x) \left[\Lop_t \rho\right](x) \,dx 
=  \int_\pS dx\, a(x)\,\int_\pS dx_0 \, \delta\left(x-f^t(x_0)\right)\rho(x_0)
\,,
\label{expLdag}
\end{equation}
meaning that one can shift the action of the push-forward operator $\Lop^\dagger_t$ from the field $a$ to the density $\rho$, that is instead pulled back by the adjoint $\Lop_t$ of the Koopman operator, called Perron-Frobenius operator.

The theory that follows is in principle valid for any chaotic dynamical system in either continuous or discrete time. On the other hand, there are specific assumptions on the spectrum of its transfer operators, namely a double spectral gap and smooth first and second eigenfunctions, as stated in the next section.  

\subsection{Chaos and spectral expansions}

The Perron-Frobenius operator $\Lop_t$ is a formal
solution to the Liouville equation $\partial_t \rho~+~\nabla~\cdot~(\rho\,v)~=~0$, where
$\dot{x} = v(x)$ is the dynamical system generating the flow $f^t(x)$. In chaotic
systems with no escape and with a proper choice of function space~\cite{BKL02}, the Perron-Frobenius
spectrum has an isolated, unit eigenvalue, whose (`leading') eigenfunction is called natural measure or invariant density~\cite{Gaspard}. The natural
measure is the weight to every phase-space average, and, as
such, its successful determination enables us to evaluate any
long-term averaged observable under the ergodicity assumption. 
For an open system, the leading eigenvalue has modulus less than one, 
and its negative logarithm is the escape rate~\cite{AltLeak}, whereas the 
associate eigenfunction is called conditionally invariant density.  
The dynamics of interest here is chaotic, so that one can choose a basis function set and a support
for $\Lop_t$ such that the resulting eigenspectrum be discrete away from the origin of the complex plane~\cite{Froy13}. For the present theory to apply, a doubly-gapped spectrum is
needed. That is not a wild assumption, 
if the system at hand is both chaotic and mixing, where time-correlations decay exponentially fast, and the near-origin continuous spectrum (within the `essential radius') is negligible, whereas the subleading part of the spectrum of $\Lop_t$ is discrete and lies within the unit circle of the complex plane~\cite{dasBuch}: in particular, the second eigenvalue yields the decay rate of an initial density to the natural measure, and it pertains an intermediate timescale,
where the system is still in the process of relaxation towards either statistical equilibrium or a stationary state.   

The meaning and avail of the second eigenfunction is the object of the present study, and it is revealed by
expanding the piece $\left[\Lop_t \rho\right](x)$ in Eq.~(\ref{expLdag}) in terms of the eigenspectrum of
$\Lop_t$:
 \begin{equation}
\int_\pS a(x) \left[\Lop_t \rho\right](x) \,dx \, =
\sum_jb_j\,e^{-\gamma_j t}\int_\pS a(x) \phi_j(x) \,dx
\label{avexpnd}
 \end{equation}
 is still the numerator of the pushed-forward average $\langle a^t\rangle$ in Eq.~(\ref{expf}), that depends on time and
 initial conditions, and thus it is not so meaningful before the system has reached statistical equilibrium.
 For that reason, it is sensible to go beyond the single moments and work out an expression for the 
 full phase-space  (`field')  distribution of $a^t$ that involves the spectrum of $\Lop_t$, which, importantly, does not
 depend on the observable chosen. Such feat can be achieved by first identifying the initial conditions in  equation~(\ref{avexpnd}) with the coefficients
 \begin{equation}
 b_j = \int_\pS \rho(x) \varphi_j(x) \, dx
 \,,
 \label{coeff}
 \end{equation}  
  where the $\varphi_j(x)$ are eigenfunctions of the Koopman operator $\Lop^\dagger_t$ (note the difference with the eigenfunctions $\phi_j(x)$ of $\Lop_t$), and then by 
  setting the initial density to $\rho(x)=\delta(x-x_0)$, that is entirely concentrated in one point.
 
 That done, the expectation value~(\ref{expf}) becomes the distribution
 \begin{equation}
\langle a^t \rangle (x_0) 
=  \frac{1}{\mu_\pS(\rho^t)}\sum_j  \varphi_j(x_0)  \,e^{-\gamma_j t}\int_\pS a(x) \phi_j(x) \,dx 
\label{distaf}
\,.
\end{equation}   
 On the other hand, using the definition of the transfer operators, as in Eq.~(\ref{expLdag}),
 \begin{equation}
 \hat{a}^{t}(x_0) :=  \langle a^t \rangle (x_0) =
   \frac{ \int_\pS dx \, a(x) \delta(x-f^t(y)) \int_\pS  dy \, \delta(y-x_0)}{\mu_\pS(\rho^t)} = 
  \frac{a(f^t(x_0))}{\mu_\pS(\rho^t)}
   \,,
   \label{at}
 \end{equation}
 which means that the dynamical average~(\ref{distaf}) evaluated locally is nothing but the 
 pushed-forward field at $x_0$, normalized by the measure of the density in the phase space. 
 The expansion~(\ref{distaf})  will be truncated to the first two terms when considering an intermediate time scale, determined by the spectral gap, as $(\gamma_1-\gamma_0)^{-1}$.
 On the other hand, the denominator $\mu_\pS(\rho^t)=\int_\pS dx\left[\Lop_t \rho\right](x)$ is approximated with the first term in the same expansion, 
 so as to obtain, overall,
 \begin{equation}
\hat{a}^{t}(x_0) 
\simeq  \int_\pS a(x) \phi_0(x) \,dx   +  \frac{\varphi_1(x_0)}{\varphi_0(x_0)}e^{-(\gamma_1-\gamma_0)t} \int_\pS a(x) \phi_1(x) \,dx\,,
\label{distrunc}
\end{equation}   
 where the first term is the long-time average weighed by the natural measure $\phi_0(x)$ (assumed normalized), while the 
 second term characterizes the \textit{profile} of the distribution by the ratio of the second to the first eigenfunctions of the Koopman operator $\Lop^\dagger_t$, that is \textit{observable independent}. If the system is closed, the escape rate vanishes and thus $\gamma_0=0$, while the measure $\mu_\pS(\rho^t)$ of the density is time independent and may be normalized to unity. That simplifies the approximate distribution~(\ref{distrunc}),  
 that now solely depends on the second eigenfunction $\varphi_1(x_0)$ of the Koopman operator for any observable.

While in dynamical averages such as Eq.~(\ref{expLdag}) it is customary~\cite{Koopmanism} to either apply
the Koopman operator to the right to the observable as $\left[\Lop^\dagger_t a\right](x)$, 
or to the left to the density as  $\left[\Lop_t \rho\right](x)$  (as Perron-Frobenius),
the converse is also possible, with the following meaning~\cite{LasMac}.   
The Perron-Frobenius operator $\Lop_t$ carries a density $\rho$, supported on $\pS$, forward in time 
to a density supported on a subset of $f^t(\pS)$. Applying the adjoint operator as
$\left[\Lop^\dagger_t \rho\right](x) = \rho(f^t(x))$, instead results in a density supported
on $f^{-t}(\pS)$, which is equivalent to go backward in time.
The corresponding adoint operations push the observable $a(x)$ forward with $\Lop^\dagger_t$, or
pull it back (`upstream') by means of the Perron-Frobenius operator, $\left[\Lop_t a\right](x)$.  

In the latter case, the backward analog of the pushed-forward field~(\ref{at}) is
 \begin{equation}
\hat{a}^{-t}(x_0) := \langle a^{-t} \rangle (x_0) =
\frac{ \int_\pS dx \, a(x) \delta(y-f^t(x)) \int_\pS  dy \, \delta(y-x_0)}{\mu_\pS(\rho^{-t})}
    = 
    \frac{a(f^{-t} (x_0))}{\mu_\pS(\rho^{-t})|J^t(f^{-t}(x_0))|} 
   \,,
   \label{amint}
  \end{equation}
  where $\rho^{-t}=\Lop_t^\dagger\rho$.  
The derivation leading to Eq.~(\ref{distrunc})
still holds under the same assumptions, but the distribution $\langle a^{-t} \rangle(x_0)$ now depends on the first two eigenfunctions of the Perron-Frobenius operator $\Lop_t$ as
 \begin{equation}
\hat{a}^{-t}(x_0)   
\simeq  \int_\pS a(x) \varphi_0(x) \,dx \,   +
\frac{\phi_1(x_0)}{\phi_0(x_0)}e^{-(\gamma_1-\gamma_0)t} \int_\pS a(x) \varphi_1(x) \,dx\,,
\label{adjdistrunc}
\end{equation} 
assuming $\varphi_0(x)$ as normalized.
Here $J^t(x)$ denotes the Jacobian matrix of the trajectory originating at $x$ and running for time $t$ (definition below), which pops up from the direct application $\left[\Lop_t\,a\right]$ of the Perron-Frobenius operator to the observable (\textit{cf.} Eq.~(\ref{expLdag})).

The universality in the profiles of the finite-time fields stems from the first-order truncation of a spectral expansion, which leads to Eqs.~(\ref{distrunc}) and~(\ref{adjdistrunc}). The timescale of validity of such approximation, and in particular the relative magnitudes of the second vs. third (neglected) term in the expansions are discussed in Appendix~\ref{AppTrunc}.

\subsection{Integrated observables} Of special interest in dynamics are the integrated observables~\cite{dasBuch}, related to Lagrangian averages~\cite{ChaotAdv17} or
finite-time Birkhoff averages (by adding a factor of $1/t$):
\begin{equation} 
A^t(x)= \int_0^t a\left[f^\tau(x)\right]\,d\tau
\,,
\label{intobs}
\end{equation}
where the integral is time ordered, that is, taken along the trajectory $f^t(x)$. The computation of $A^t(x)$ is straightforward if
the observable $a(x)$ is a functional field, for example  in the phase-space diffusivity
\begin{equation} 
D^t(x)= \int_0^t \left[f^\tau(x)-x\right]^2\,d\tau
\,,
\end{equation}
but more involved if $a(x)$ is an operator, as it is the case for the Jacobian    
\begin{equation}
J^t(x_0) = e^{\int_0^t d\tau M(f^\tau(x_0))}
\,, \hskip 0.5cm M_{ij}(x) = \frac{\partial v_i(x)}{\partial x_j}
\,,
\label{Jac}    
\end{equation}
where the exponential is the formal solution to the matrix differential equation $\frac{dJ^t}{dx}=MJ$ (sometimes called variational initial-value problem~\cite{Dankowicz}) along the orbit $x_0\rightarrow f^t(x_0)$.
The quantity of interest is then
the stability (or finite-time Lyapunov) exponent $A^t(x_0) = \ln ||J^t(x_0)||/t$.

The expectation value~(\ref{expf}) can now be generalized to an integrated observable
 $A^t(x)$ by either $i)$ choosing the mapped $\left[\Lop_t \rho\right](x)$ as weight function
\begin{equation}
\langle A^t\rangle_{\rho^t}  = \int_\pS A^t(x) \left[\Lop_t \rho\right](x) \,dx 
\,,
\end{equation}
which leads to the approximate field distribution pinned at the initial point of the trajectory 
$x_0\rightarrow f^t(x_0)$:
 \begin{equation}
 \hat{A}^t(x_0) :=
\frac{A^t (x_0)}{\mu_\pS(\rho^t)} \simeq  \int_\pS A^t(x) \phi_0(x) \,dx   +  \frac{\varphi_1(x_0)}{\varphi_0(x_0)}e^{-(\gamma_1-\gamma_0)t} \int_\pS A^t(x) \phi_1(x) \,dx\,,
\label{Adistrunc}
\end{equation}  
or $ii)$ by choosing the initial $\rho(x)$ as weight function
 \begin{equation}
\langle A^t\rangle_{\rho}  = \int_\pS A^t(x) \rho(x) \,dx 
\,,
\end{equation}
that results in the observable field pinned at the arrival point of the trajectory  $f^{-t}(x_0)\rightarrow x_0$:
 \begin{equation}
 \hat{A}^{-t}(x_0) :=
\frac{A(f^{-t} (x_0))}{\mu_\pS(\rho^{-t})|J^t(f^{-t}(x_0))|} \simeq  \int_\pS A^t(x) \varphi_0(x) \,dx \,   +
\frac{\phi_1(x_0)}{\phi_0(x_0)}e^{-(\gamma_1-\gamma_0)t} \int_\pS A^t(x) \varphi_1(x) \,dx\,.
\label{Aadjdistrunc}
\end{equation} 
A full account of time-forward and backward integrated observables for chaotic relaxation processes has been given elsewhere~\cite{LippTherm}. Here, it should be remarked that in both expressions~(\ref{Adistrunc}) and~(\ref{Aadjdistrunc}) the time-dependence of the integrals $\int_\pS A^t(x) \varphi_i(x)$ (or  $\int_\pS A^t(x) \phi_i(x)$ backward) can be written explicitly, if the observable $a(x)$ is a simple function of the phase-space coordinates. In that case (see Appendix~\ref{AppTrunc} for details), the distribution of a forward integrated observable~(\ref{Adistrunc}) (and likewise the backward integration~(\ref{Aadjdistrunc}))  becomes
 \begin{equation}
 \hat{A}^t(x_0) 
\simeq  \frac{1-e^{-\gamma_0 t}}{\gamma_0}\int_\pS a(x) \phi_0(x) \,dx   +   e^{-(\gamma_1-\gamma_0)t}
\frac{1-e^{-\gamma_1 t}}{\gamma_1}
\frac{\varphi_1(x_0)}{\varphi_0(x_0)} \int_\pS a(x) \phi_1(x) \,dx\,,
\label{AdistruncFunc} 
\end{equation}
which produces a non-monotonic decay for the second term, and a longer relaxation time to equilibrium or stationarity than for local observables. This aspect will also be featured in the numerical analysis of section~\ref{numerics}.  

\subsection{Noise}

The above analysis and predictions may be generalized to chaotic systems subject to weak noise, so that
the equations of motion take the Langevin form $\dot{x} = v(x) + \eta(t)$, with the time-uncorrelated and  Gaussian-distributed random force $\eta(t)$. The transfer operator $\Lop_t$ ($\Lop_t^\dagger$), that is
the formal solution to the associate Fokker-Planck equation~\cite{Risken}, features a noisy kernel, for instance~\cite{CviLip12} 
\begin{equation}
\Lop_{\Delta,t}\,\rho(x) = \frac{1}{\sqrt{4\pi \Delta t}}\int_\pS dx\, e^{-(y-f^t(x))^2/4\Delta t}\rho(x)
 \,,
 \label{FPevol}
 \end{equation} 
for isotropic noise  of amplitude $2\Delta$. The operator $\Lop_{\Delta,t}$ and its adjoint can be applied to a density $\rho$ as above, or to an observable $a$, to obtain its noisy evolution, whereas the integrated observable~(\ref{intobs}) takes the form
\begin{equation} 
A^t_\Delta(x)=\int_0^t \left[\Lop^\dagger_{\Delta,\tau}a\right](x)\,d\tau
\,.
\label{intobsns}
\end{equation}

\section{Validation}
\label{numerics}
The above predictions are now tested with three different models of chaos, the Bernoulli map, the perturbed cat map and 
the Hamiltonian H\'enon map. 
While the first system is treated analytically, the other two are tackled numerically.
For all models, the ratio of the first two eigenfunctions of the Perron-Frobenius and of the Koopman spectrum is compared with the distributions of different pushed-forward, pulled-back, or integrated
observables for finite time. 

\subsection{Bernoulli map}
\label{BerSec}
It is defined as 
\begin{equation}
 f(x) = 2x\,\mathrm{mod}\,1 = \left\{
  \begin{array}{c} 
  2x \hspace{0.9cm} 0\leq x< \frac{1}{2}  \\   
  2x -1 \hspace{0.5cm} \frac{1}{2}\leq x< 1  
\,.
\end{array}
\right.
\label{BerMap}
\end{equation}
This one-dimensional, non-invertible map, exhibits chaos everywhere on the unit interval, admits no escape, and it has a constant Lyapunov exponent equal to $\ln 2$.

The Perron-Frobenius operator acts on a density
at each time step as
\begin{equation}
\Lop\rho(x) = 	\frac{1}{2}\left[ \rho\left(\frac{x}{2}\right) +  \rho\left(\frac{x+1}{2}\right)\right]
\,,
\label{BerPF}
\end{equation}
and it has been spectrally decomposed in $L^2$~\cite{Driebe} by means of  
the Bernoulli polynomials 
\begin{eqnarray}
\phi_0(x) &=& 1 \\
 \phi_1(x) &=& x - \frac{1}{2} \label{BernSec}\\
 \phi_2 (x) &=& x^2 -x + \frac{1}{6}  \label{BerPFEig} \\
   &\ldots& \label{BernEtc}
\end{eqnarray}
as eigenfunctions for $\Lop$ of eigenvalues $\lambda_n=2^{-n}$. The one-dimensional nature of the phase space makes the forward action~(\ref{BerPF}) all expanding, while the backward, Koopman operator~\cite{Fox}
\begin{equation} 
\Lop^\dagger\rho(x) = 	\rho(2x)\Theta\left(\frac{1}{2}-x\right) +  \rho(2x-1)\Theta\left(x-\frac{1}{2}\right)
\,,
\label{BerKoop}
\end{equation}
is everywhere squeezing (here $\Theta$ is the Heaviside step function). Its leading eigenfunction $\varphi_0(x)=1$ is again uniform on the unit interval, while the rest of the spectrum is made of the generalized functions
\begin{equation}
\varphi_j(x) = \frac{(-1)^{j-1}}{j!}\left[\delta_-^{j-1}(x-1) -\delta_+^{j-1}(x)\right]
\,,  
\label{BerKoopEig}
\end{equation}
for $j\geq1$, that is combinations of  Dirac delta functions and their derivatives. This behavior is 
peculiar of one-dimensional chaotic maps. Due to the result~(\ref{BerKoopEig}),  we may not apply the expansion~(\ref{distrunc})~[(\ref{Adistrunc})], that yields a local [integrated] observable $\hat{a}^t(x)$ [$\hat{A}^t(x)$] in terms of the 
first two eigenfunctions $\varphi_0$ and $\varphi_1$ of the Koopman operator,  since  the present theory assumes the $\varphi_i$'s to be smooth. Instead, we may test the validity of the 
expressions~(\ref{adjdistrunc}) and (\ref{Aadjdistrunc}), that relate the polynomial eigenfunctions~(\ref{BernEtc}) to the      
observable $\hat{a}^{-t}(x)$ or $\hat{A}^{-t}(x)$, pinned by the final points of the orbits $f^{-t}(x)\rightarrow x$.

Two test observables are considered here, namely $a_1(x)=x^2$, and $a_2(x)=x+\sin20x$, which the Perron-Frobenius operator~(\ref{BerPF}) is applied to for a time $t$ before the system relaxes to equilibrium. 
\begin{figure}[tbh!]
\centerline{
(a)\scalebox{.3}{\includegraphics{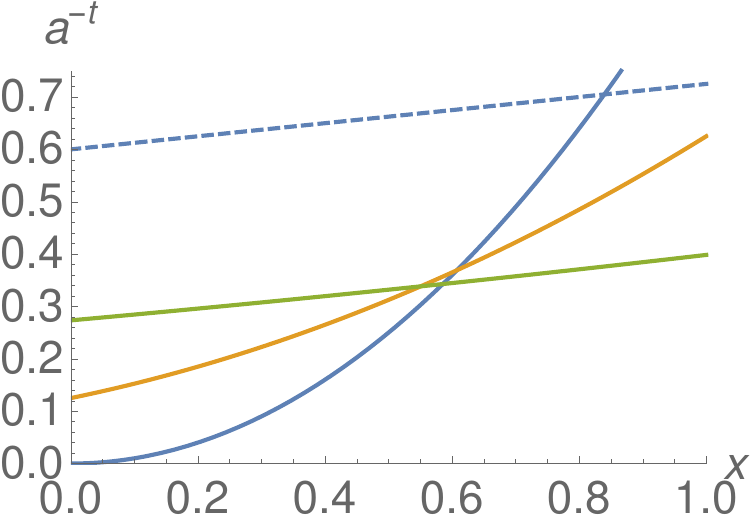}}
(b)\scalebox{.3}{\includegraphics{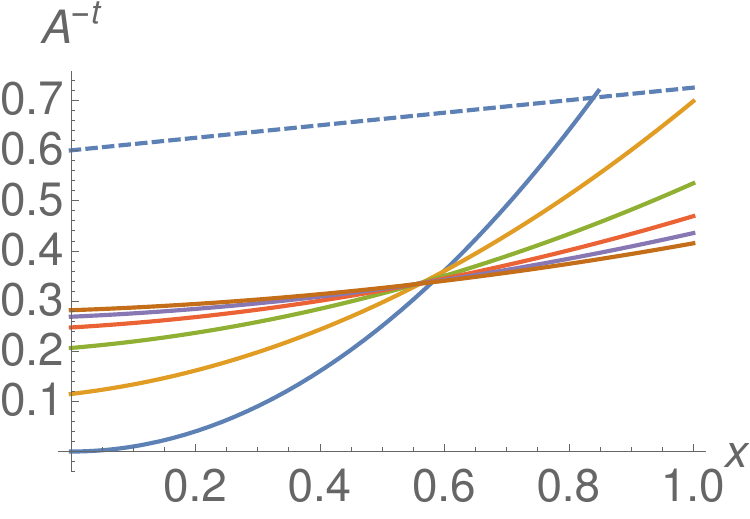}}
(c)\scalebox{.3}{\includegraphics{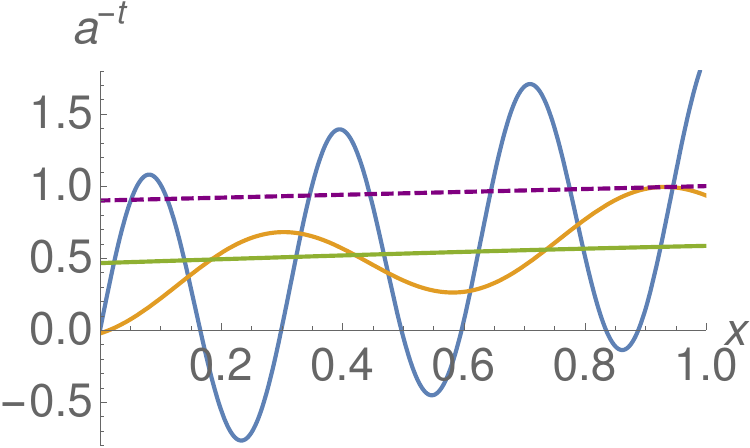}}
(d)\scalebox{.3}{\includegraphics{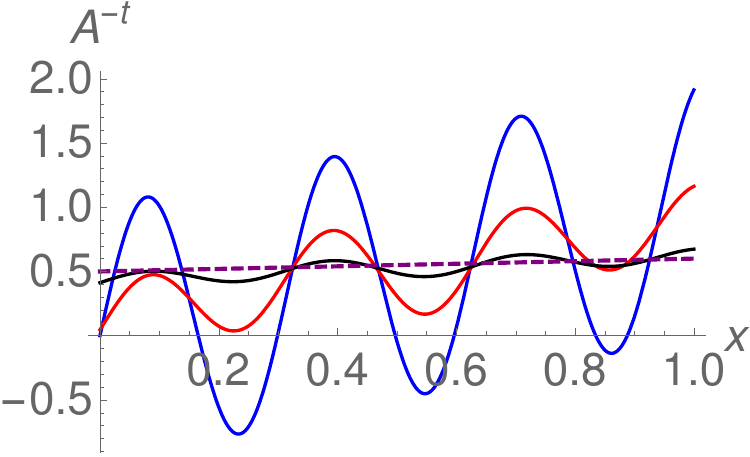}}}
\caption{Test observables mapped by the Perron-Frobenius operator and time integrated: (a) $a_1=x^2$ (blue, solid line) and successive $\hat{a}^{-t}$ (solid lines in color) with $t=2$ and $t=4$,  approaching a straight line (for increasing $t$); (b)
 $\hat{A}^{-t}_1(x)$ (solid lines in color)
with $t=3$, $t=5$, $t=8$, $t=10$, $t=15$ approaching a straight line (for increasing $t$), plotted above (dashed line) for comparison; (c) $a_2 = x +\sin20x$ (blue, solid line), and successive  $\hat{a}^{-t}$
(solid lines in color) with $t=2$ and $t=5$, the latter being practically a straight line; (d)
$\hat{A}^{-t}_2(x)$ (solid lines in color) with $t=3$, $t=15$ approaching a straight line (dashed line).}
\label{BerObs}    
 \end{figure}
Applying $\Lop_t$ has the effect of progressively smoothing observables as $t$ increases, e.g.  
\begin{equation}
[\Lop_{t=5} a_1](x) = 0.3 + 0.06x+ 0.004x^2
\,,
\end{equation}
or
\begin{equation} 
[\Lop_{t=5} a_2](x) \simeq 0.46 + 0.14x+ 0.002x^2
\,,
\end{equation}
when expanding the closed-form (but cumbersome) expression of $[\Lop_{t=5} a_2](x)$ in a power series and neglecting the cubic- and higher-order terms. As a consequence, the mapping of both $a_1$, $a_2$ to $\hat{a}_1^{-t}$, $\hat{a}_2^{-t}$  produces curves that tend to a straight line, and, at a longer timescale, the corresponding integrated observables $\hat{A}_1^{-t}$, $\hat{A}_2^{-t}$, also approach the functional form of the second eigenfunction~(\ref{BernSec}) of the 
Perron-Frobenius spectrum (Fig.~\ref{BerObs}),  $\phi_1(x) = x - \frac{1}{2}$. Specifically, Eqs.~(\ref{adjdistrunc}) and~(\ref{Aadjdistrunc}) predict that $\hat{a}^{-t}(x) \sim \phi_1(x)/\phi_0(x)$
(idem for $\hat{A}^{-t}$), where $\phi_0(x)=1$ in 
this case, while  the proportionality factor (slope of the line) depends on the observable.     

\subsection{Two-dimensional models}
First, the transfer operator is discretized by means of  Ulam's method~\cite{Ulam},
that is by subdividing the phase space into $N$ intervals
$\pS_i$ of equal area, 
and evaluating the transition probabilities from 
$\pS_i$ to $\pS_j$~\cite{ChapTan}. 
First, densities are written as sums of characteristic functions on a phase-space partition
\begin{equation}
\rho(x) = \sum_j^N \rho_j \frac{\chi_j(x)}{\mu(\pS_j)}
\,,
\end{equation}
where $\chi_j(x)=1$ for $x\in\pS_j$, and $\chi_j(x)=0$ otherwise, while the $\rho_j$'s are non-negative
components in this basis.
The action of the Perron-Frobenius operator on a density is then rewritten as
\begin{equation}
\left[\Lop_t\rho\right]_i = \sum_j^N \frac{\rho_j}{\mu(\pS_j)}\int_{\pS_i} dx \int_{\pS_j} dy\, \delta(x-f^t(y))
= 
\sum_j^N \frac{\mu\left(\pS_j \bigcap f^t(\pS_i)\right)}{\mu(\pS_i)} \rho_j 
\,,
\end{equation}
which yields the Ulam matrix
\begin{equation}
[\mathbf{L}_t]_{ij} = \frac{\mu\left(\pS_j \bigcap f^t(\pS_i)\right)}{\mu(\pS_i)}
\,.
\label{Ulmat}
\end{equation}
The numerator of the transition rate~(\ref{Ulmat}) is estimated with a  
Monte Carlo method~\cite{ErmShep}, that consists of iterating random initial conditions from each cell $\pS_i$ 
and counting which fraction lands in each $\pS_j$. The Ulam matrix acts to its right as the Perron-Frobenius operator, and to its left as its adjoint, namely the Koopman operator.   

The finite-dimensional approximation $\mathbf{L}_t$ obtained with the Ulam method leverages the Markov property of the systems of our interest, but it is in general unstable
and must be used with caution~\cite{Froy07}. Still, for certain everywhere unstable (`hyperbolic') maps,
$\mathbf{L}_t$  was found to well reproduce the leading- and first subleading eigenfunctions
of the Perron-Frobenius operator on a space of characteristic functions supported on the intervals $\pS_i$ at sufficiently large $N$~\cite{Kensuke,ClassScars}.

\subsubsection{Perturbed cat map}
\label{catsec}
It is defined by $f(x) = T_\epsilon \circ T[x]$, with $x=(q,p)$,
 \begin{equation}
 T  \left( \begin{array}{c}
 q \\
 p
 \end{array} 
 \right)= \left(\begin{array}{cc}
 1 & 1 \\
 1 & 2 
 \end{array}
 \right)
  \left( \begin{array}{c}
 q \\
 p 
 \end{array}
 \right) \hskip0.2cm
 \mathrm{mod} 1
 \,,
 \end{equation}
 and
 \begin{equation}
 T_\epsilon 
 \left( \begin{array}{c}
 q \\
 p 
 \end{array}
 \right)
 = \left(\begin{array}{c}
 q - \epsilon\sin2\pi p \\
 p
 \end{array}
 \right) \hskip0.2cm
 \mathrm{mod} 1
 \,,
 \end{equation}
with $\epsilon=0.1$ throughout the simulations.
This system is strongly chaotic and hyperbolic, that is, correlations decay exponentially fast with time~\cite{Arnold}. It possesses an infinite number of unstable periodic orbits, 
and, specifically, a fixed point at the origin. The phase space is a 2-torus, there is no escape, and areas are preserved by the time evolution, so that the determinant of the Jacobian 
matrix of every trajectory is equal to unity.

It is noted that the Perron-Frobenius operator is unitary for area-preserving maps acting on
$L^2$. As a consequence, the spectrum should lie entirely on the unit circle of the complex plane, and
there would be no relaxation to the natural measure. However, the discretization~(\ref{Ulmat}) used here to compute the spectrum, effectively introduces noise of amplitude
proportional to the mesh, which breaks unitarity and opens a spectral gap~\cite{Kensuke}.
The eigenvalues of the Perron-Frobenius operator then tend to the Pollicott-Ruelle resonances
in the noiseless limit~\cite{FaureRoy}.     

Because the perturbed cat map has no escape and it is area preserving,  the first eigenfunction of both the Perron-Frobenius and the Koopman operators is a uniform distribution. 
Thus, according to equations~(\ref{distrunc}) and~(\ref{adjdistrunc}), the out-of-equilibrium distribution of any pushed-forward, pulled-back or integrated observable is characterized by the second eigenfunction alone.    
 \begin{figure}[tbh!]
\centerline{
(a)\scalebox{.35}{\includegraphics{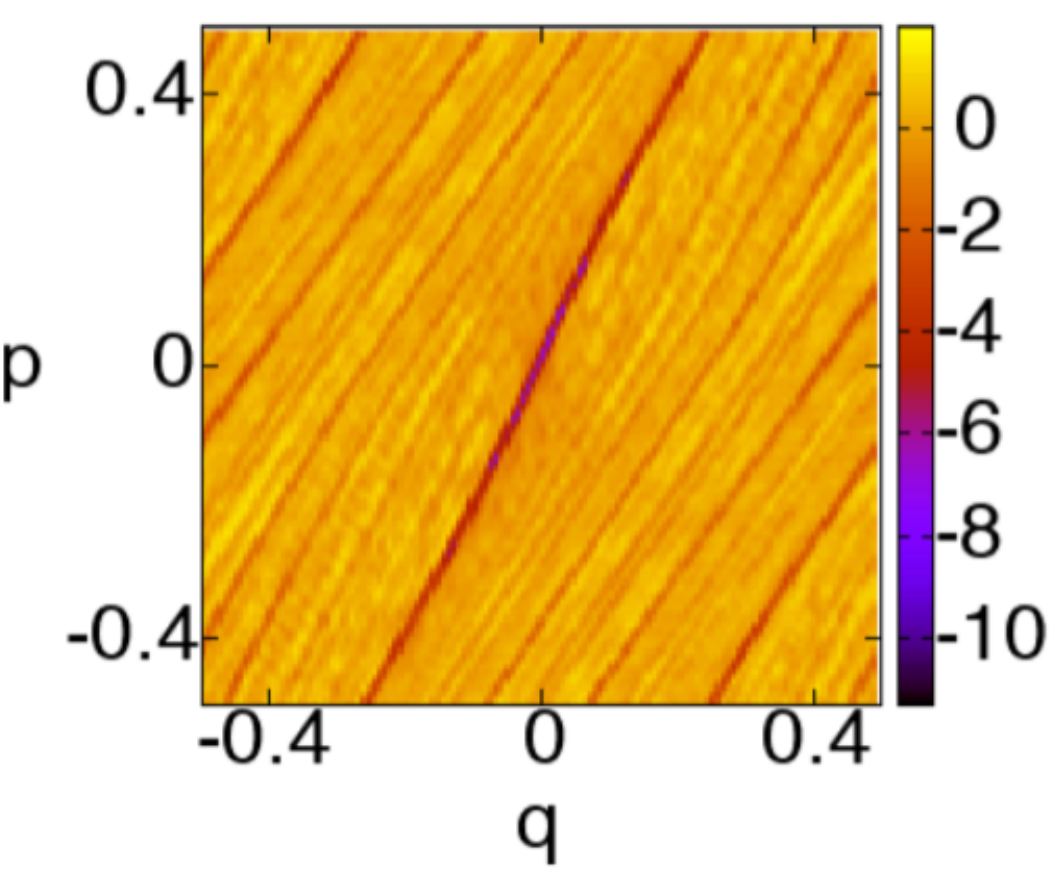}}
(b)\scalebox{.35}{\includegraphics{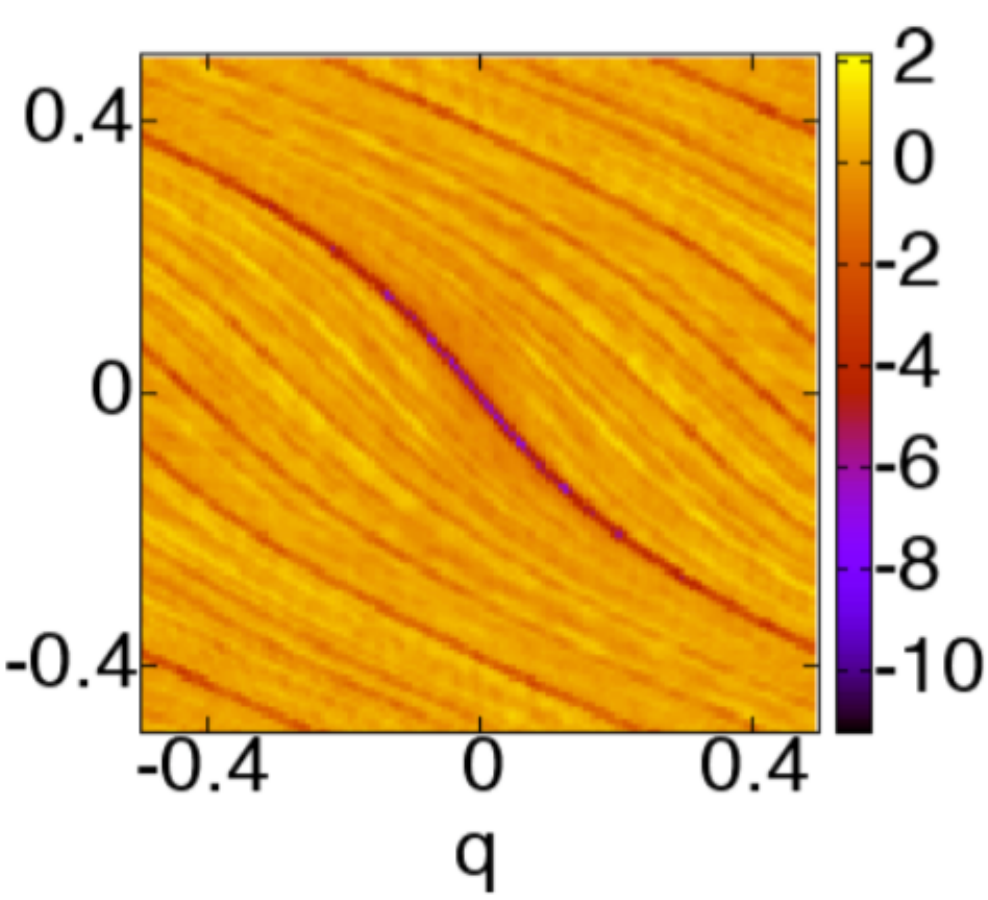}}}
\caption{First subleading eigenfunctions of the (a) Perron-Frobenius- and (b) Koopman operator for the perturbed cat map. The Ulam matrix has size $2^{14}\times2^{14}$.}
\label{Sublead}
\end{figure}

Figure~\ref{Sublead} shows the numerically computed second eigenfunctions of the Perron-Frobenius and Koopman operators with the Ulam discretization. They are striated along the unstable and 
stable manifolds respectively, that emanate from the fixed point at the origin.  
\begin{figure*}[tbh!]
\centerline{
(a)\scalebox{.35}{\includegraphics{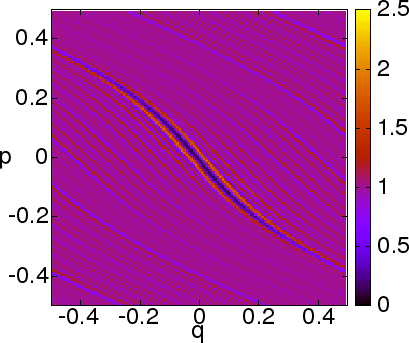}}
(b)\scalebox{.35}{\includegraphics{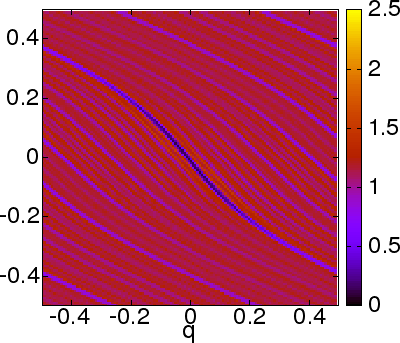}}
(c)\scalebox{.35}{\includegraphics{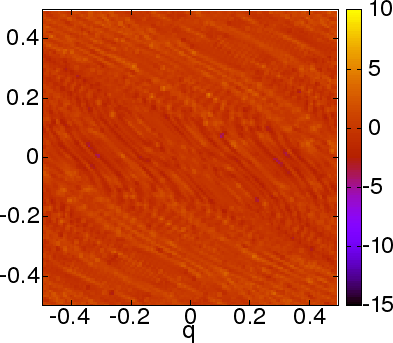}}}
\vskip 0.2cm
\centerline{
(d)\scalebox{.45}{\includegraphics{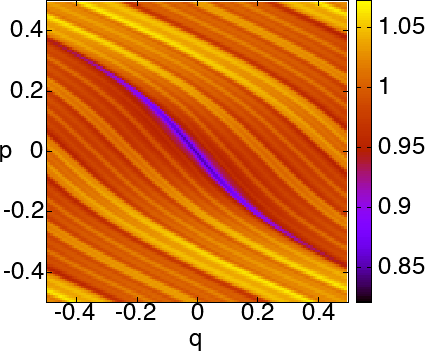}}
(e)\scalebox{.45}{\includegraphics{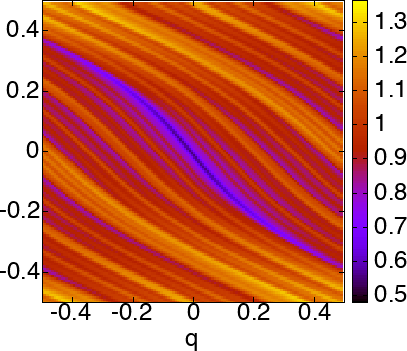}}
(f)\scalebox{.47}{\includegraphics{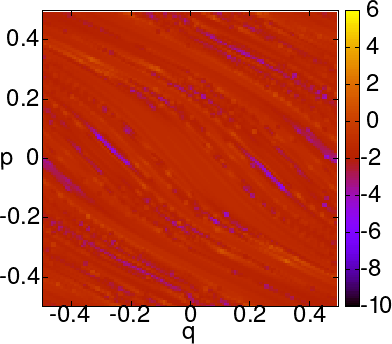}}}
\caption{Perturbed cat map: the phase-space distribution  ($2^{14}$ points, each averaged over $10^4$ trajectories) of (a) the local kinetic energy  $\frac{1}{2} p^2(f^t(x))$ (t=6) vs. (b) the observable $q^2(f^t(x))$ (t=7), and (c) the logarithmic ratio~(\ref{ratio}) between the two. (d)   
 The field distributions of  the finite-time Lyapunov exponents  (t=15), and (e) the integrated kinetic energy (t=15), versus $x=(q,p)$, initial point of the trajectory $x\rightarrow f^{t}(x)$. (f) 
  The logarithmic ratio~(\ref{ratio}) between the two.}
\label{CatInitt5}
\end{figure*}

\begin{figure*}[tbh!]
\centerline{
(a)\scalebox{.35}{\includegraphics{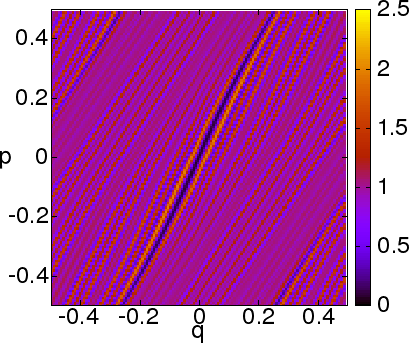}}
(b)\scalebox{.35}{\includegraphics{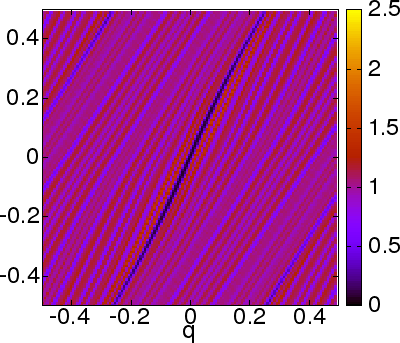}}
(c)\scalebox{.35}{\includegraphics{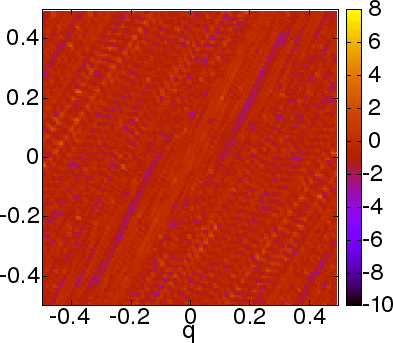}}}
\vskip 0.2cm
\centerline{
(d)\scalebox{.45}{\includegraphics{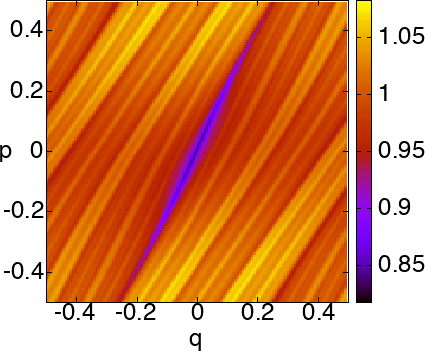}}
(e)\scalebox{.45}{\includegraphics{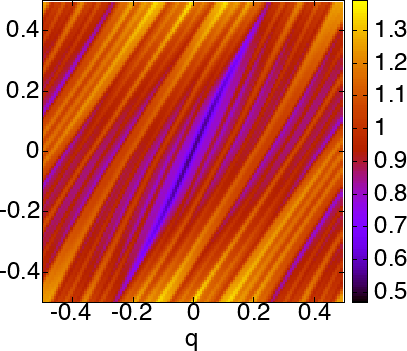}}
(f)\scalebox{.47}{\includegraphics{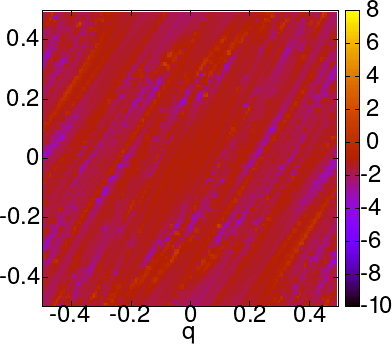}}}
\caption{Perturbed cat map: the phase-space distribution  ($2^{14}$ points, each averaged over $10^4$ trajectories) of (a) the local kinetic energy  $\frac{1}{2} p^2(f^{-t}(x))$ (t=6) vs. (b) the observable $q^2(f^{-t}(x))$ (t=6), and (c) the logarithmic ratio~(\ref{ratio}) between the two. (d)   
 The field distributions of  the finite-time Lyapunov exponents  (t=15), and (e) the integrated kinetic energy (t=15), versus $x=(q,p)$, final point of the trajectory $f^{-t}(x)\rightarrow x$. (f) 
  The logarithmic ratio~(\ref{ratio}) between the two.}
\label{CatFint5}
\end{figure*}

On the other hand, figures~\ref{CatInitt5}-\ref{CatFint5} illustrate the behavior of the numerically computed finite-time fields of three distinct observables, namely the stability (Lyapunov) exponent stemming from the Jacobian~(\ref{Jac}),
the time-integrated kinetic energy $\overline{K}^t(x) = \frac{1}{2}\int_0^t d\tau p^2(f^\tau(x))$, and the local observables $q^2(x)$ and $p^2(x)$, which are both pushed forward as $q^2(f^t(x))$ or  $p^2(f^t(x))$  , and pulled back as
$q^{2}(f^{-t}(x))$ or $p^{2}(f^{-t}(x))$. The fields portrayed in the figures are obtained by iterating some $10^8$ randomly chosen, uniformly distributed initial conditions until a certain time $t$ before relaxation.  
Distinct local and integrated observables exhibit nearly identical profiles, within the timescales considered.
When pushed forward or pinned by the initial points $x_0$ of the trajectory $x_0\rightarrow f^t(x_0)$ (figure~\ref{CatInitt5}(a)(b),(d)(e)), 
all finite-time phase space distributions share the philamentary pattern of the second eigenfunction of the Koopman operator (figure~\ref{Sublead}(b)), marked by a continuous direction along the stable manifold of the map, and a fractal-like cross section.  
That is predicted by equations~(\ref{distrunc}) and~(\ref{Adistrunc}).
Similarly, chaotic fields of both observables follow the 
behavior of the second eigenfunction of the Perron-Frobenius operator (figure~\ref{Sublead}(a)) when pinned by the final point of the same trajectory (figure~\ref{CatFint5}(a)(b),(d)(e)),
as expected from equations~(\ref{adjdistrunc}) and~(\ref{Aadjdistrunc}). 
This time the striation goes along the unstable manifold of the map.

In addition, 
the logarithmic ratio
\begin{equation}
 r(x_0,t) = \ln \left|\frac{\hat{a}_1^t(x_0) - \langle a^t_1 \rangle_\pS}
 {\hat{a}_2^t(x_0) - \langle a^t_2 \rangle_\pS}\right|
 \label{ratio}
 \end{equation}
 with $\langle a \rangle_\pS =  \int_\pS a(x) \phi_0(x) \,dx$,  
 is reported in figures~\ref{CatInitt5}-~\ref{CatFint5}(c)(f), in order to compare the profiles of the distributions $\hat{a}_1^t(x_0)$ and $\hat{a}_2^t(x_0)$ of two distinct 
 observables, and quantify their differences as the deviation of equation~(\ref{ratio}) from uniformity.
 In fact, the theory developed here predicts that fields should follow the second eigenfunction 
 of the transfer operator up to an additive and a multiplicative constants, both depending on the 
 observable. For observable-independent profiles, the quantity  $r(x_0,t)$ is uniform in the phase space,
 up to fluctuations, otherwise one would expect $r(x_0,t)$ to feature patterns aligned with the eigenfunctions along the (un)stable manifold. 
  The density plots in figures~\ref{CatInitt5}-~\ref{CatFint5}(c)(f) do not show any such patterns, altghough erratic striations can be recognized at rougher scales than those of the field plots.   
 That supports the claim of near identity of the fields profiles compared, whithin the attainable accuracy
  of the numerical observations.  

It is important to remark that significantly shorter iteration times were chosen for the local ($t\sim7$) than for the 
integrated ($t\sim15$) observable distributions, since the former relax to equilibrium faster (and monotonically)  than the latter, as discussed in Appendix~\ref{AppTrunc}.    

\subsubsection{Hamiltonian H\'enon map with noise}
The next model to test the theory on is the Hamiltonian H\'enon map
 \begin{equation}
 f  \left( \begin{array}{c}
 q \\
 p
 \end{array} 
\right) 
 = \left(\begin{array}{c}
1 -\alpha q^2 +\beta p \\
 q
 \end{array}
 \right)
\label{HamHen}
\end{equation}
 with $\alpha=1.4$ and $\beta=-1$. This map has no dissipation, but it does allow escape to infinity from the 
 neighborhood of the two fixed points: $x_p\simeq(-1.1,-1.89)$ is unstable (`hyperbolic') and generates a chaotic saddle through
 its stable and unstable manifolds, while $x_c\simeq(0.39,0.39)$ is marginally stable (`center'), and surrounded by a tiny stability island.
 In order to `kick' the dynamics out of the latter non-chaotic region and into the chaotic phase, weak noise is 
 added to the map~(\ref{HamHen}) of an amplitude comparable to the size of the stability island per unit time.
 Strictly speaking, the non-hyperbolicity of the resulting noisy system should introduce a continuous component
 in the spectrum of the transport operators and thus break the assumption of a solely discrete spectrum. However, if we investigate
 timescales of the order of- or shorter than the inverse escape rate from the chaotic saddle, when the 
 discrete part of the spectrum is dominant, the contribution of the continuous part of the spectrum
 may be ignored, due to the smallness of the stability island.
 Because the system admits escape and the first eigenfunction $\phi_0$ ($\varphi_0$) of $\Lop_t$ ($\Lop^\dagger_t$) is non-uniform, 
 the phase-space distributions of the pushed-forward, pulled-back or integrated observables for intermediate time scales are determined by the 
 ratio of the second to the first eigenfunction of the evolution operator. 
 
 The predictions~(\ref{Adistrunc}) and~(\ref{Aadjdistrunc})
 are tested for two observables, that is the finite-time stability exponent and the diffusivity $\hat{D}^t(x) = \frac{1}{t} \int^t_0 d\tau \, [q^2(f^\tau(x)) + p^2(f^\tau(x))]$. 
 \begin{figure*}[tbh!]
\centerline{
(a)\scalebox{.47}{\includegraphics{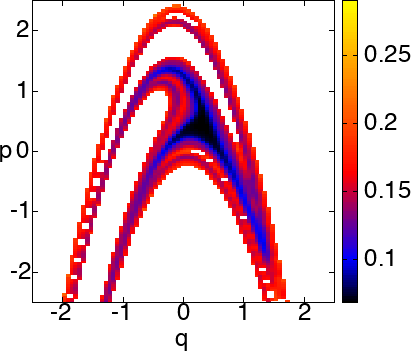}}
(b)\scalebox{.47}{\includegraphics{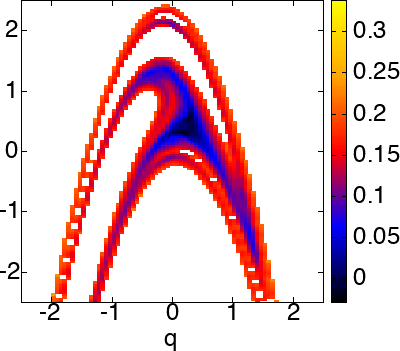}}
(c)\scalebox{.49}{\includegraphics{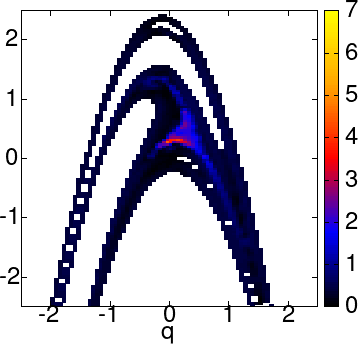}}}
\vskip 0.2cm
\centerline{
(d) \scalebox{.47}{\includegraphics{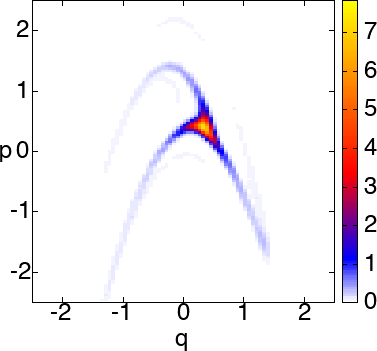}}
(e) \scalebox{.47}{\includegraphics{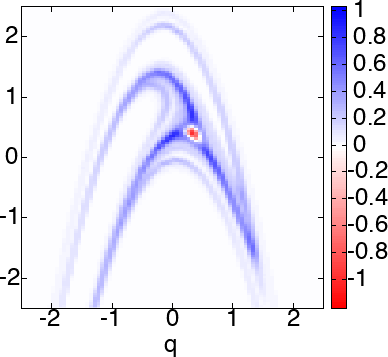}}
(f)\scalebox{.47}{\includegraphics{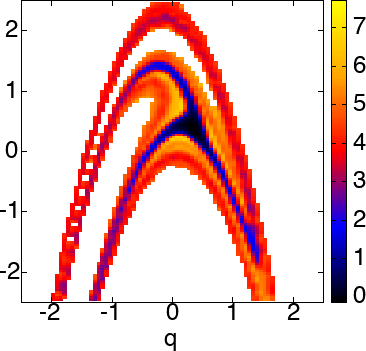}}}
\caption{Phase-space distributions ($2^{14}$ points, each averaged over $10^4$ trajectories) of integrated oservables of the H\'enon map with white, uniformly-distributed noise of amplitude $\Delta=10^{-3}$, pinned by the initial points, after $t=10$ iterations of the map: (a) diffusivity; (b): finite-time Lyapunov exponents; (c) ratio of the distributions of the observables in (a) and (b), as defined in~(\ref{ratio}).
(d) The  leading eigenfunction of the Koopman operator for the same noisy map, approximated with the Ulam method on a matrix of size $2^{14}\times2^{14}$; (e) the first subleading eigenfunction of the same matrix; (f) the ratio (in absolute value) of (e) to (d).}
\label{henobs}
\end{figure*}  
The density plots in figure~\ref{henobs}(a)(b) corroborate the expectations for the field distributions of the two integrated observables to be supported on the 
stable manifold of the map when pinned by the initial points of the iteration $x_0\rightarrow f^t(x_0)$ plus weak noise, and to mimic the density profile of the ratio between the 
second and the first eigenfunction of the Koopman operator (figure~\ref{henobs}(f)). The two distinct fields in figure~\ref{henobs}(a)(b) display nearly identical profiles, as once again 
confirmed by the nearly uniform pattern of their ratio, Eq.~(\ref{ratio}), Fig.~\ref{henobs}(c).  

\begin{figure*}[tbh!]
\centering
(a)\scalebox{.45}{\includegraphics{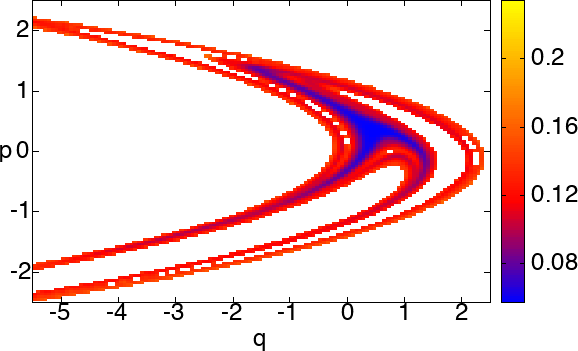}}
\hskip 0.5cm
(b)\scalebox{.45}{\includegraphics{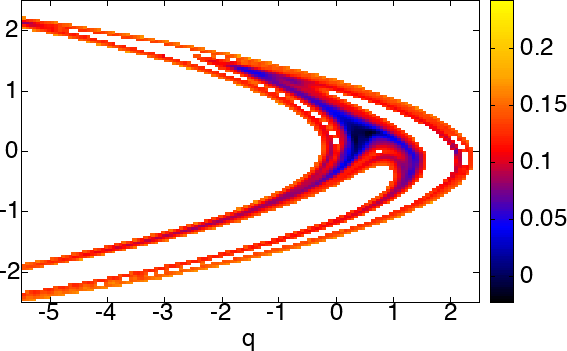}}
\vskip 0.2cm
(c) \scalebox{.45}{\includegraphics{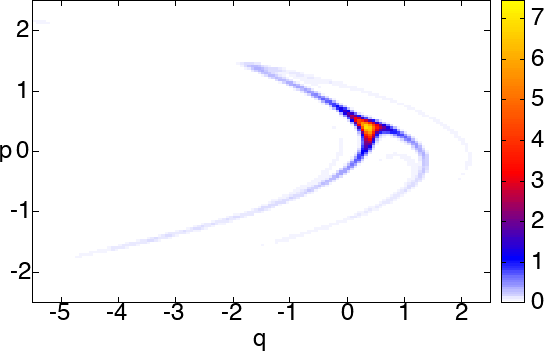}}
\hskip 0.5cm
(d) \scalebox{.45}{\includegraphics{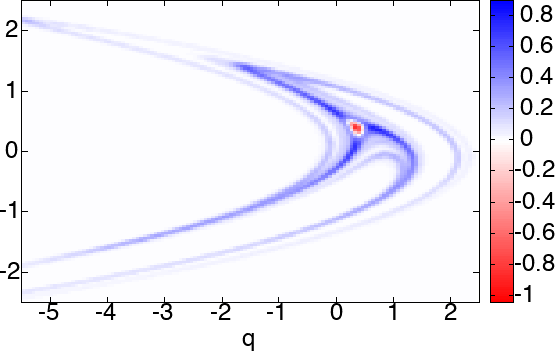}}
 \vskip 0.2cm
(e)\scalebox{.45}{\includegraphics{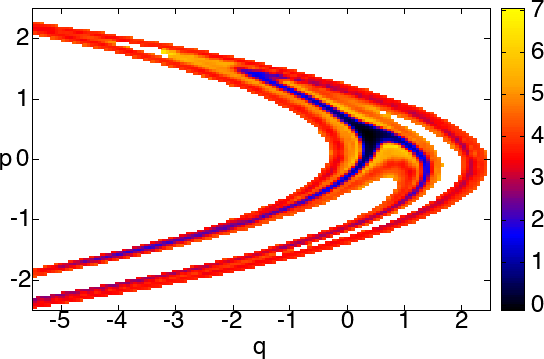}}
\hskip 0.5cm
(f)\scalebox{.47}{\includegraphics{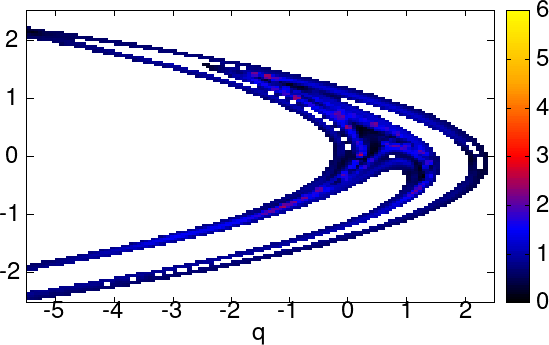}}
\caption{Phase-space distributions ($2^{14}$ points, each averaged over $10^4$ trajectories) of integrated oservables of the H\'enon map with white, uniformly-distributed noise of amplitude $\Delta=10^{-3}$, pinned by the final points, after $t=10$ iterations of the map: (a) diffusivity; (b): finite-time Lyapunov exponents. (c) The  leading eigenfunction of the Perron-Frobenius operator for the same noisy map, approximated with the Ulam method on a matrix of size $2^{14}\times2^{14}$; (d) the first subleading eigenfunction of the same matrix; (e) the ratio (in absolute value) of (d) to (c). (f) Ratio of the distributions of the observables in (a) and (b), as defined in~(\ref{ratio}).}
\label{henbobs}
\end{figure*}

Analogously, the same two observables pinned by the final points 
of each phase-space trajectory $f^{-t}(x_0)\rightarrow x_0$ plus weak noise produce fields
supported on the unstable manifold of the map (Fig.~\ref{henbobs}(a)(b)). 
Lyapunov exponent and diffusivity exhibit twin profiles (see their ratio in (Fig.~\ref{henbobs}(f)), 
and behave similarly to the ratio of the second to the first eigenfunction of the Perron-Frobenius operator (Fig.~\ref{henbobs}(e)). 

In both `forward' and `backward' pictures, the strongly chaotic phase (in orange) is distinguishable from the non-hyperbolic, weakly chaotic phase (in blue) of a three-lobed shape with tapered ends, due to a period-three unstable periodic orbit that rules the dynamics just outside the stability island. 

The white color in figures~\ref{henobs}-\ref{henbobs}(a)(b) represents the region of the phase space where forward (figure~\ref{henobs}) or backward (figure~\ref{henbobs}) trajectories escape from the domain examined before the time $t$ of integration, and thus it is not part of the field distributions. In figures~\ref{henobs}(f) and~\ref{henbobs}(e), instead, the ratio between the eigenfunctions is not defined in the blank region, where the leading eigenfunction vanishes. 
The density plots of the leading- and subleading eigenfunctions of the transport operators (figures~\ref{henobs}(d)(e) and~\ref{henbobs}(c)(d)) taken separately, bear significant differences from the fields of the observables: the first eigenfunctions clearly describe a longer timescale than that of the field distributions, at which noisy trajectories have mostly left the hyperbolic region, while they only survive in and around the stability island; the second eigenfunctions alone are more resemblant of the finite-time fields, except they are suppressed on a ring around the stability island, a feature that does not appear in the density plots
of the observables. That supports the necessity of computing the ratio between the eigenfunctions to 
reproduce the field profiles.               

\section{Discussion}
Time-dependent phase-space distributions  of observables advected by a chaotic dynamical system and before relaxation to statistical equilibrium  exhibit universal features, 
determined by the first two eigenfunctions of the Perron-Frobenius or Koopman transport operators. 

This result essentially owes to three separate intuitions throughout the derivation: first, in a dynamical average, the action of the
evolution operator can be shifted at will between the observable and the density; second, a time-dependent 
average is turned into the more meaningful phase-space distribution by setting the initial condition to a delta function centered at any point; third, the first non-trivial order truncation of the expansion of a pushed-forward or pulled-back chaotic field in terms of eigenfunctions of the transport operators gives rise to observable-independent field profiles, and thus universality at an intermediate timescale
of integration.

The theory is validated on two-dimensional models of chaos, but its implications go as far as the applicability of the Koopman treatment to problems of fluid dynamics (e.g. passive scalar turbulence~\cite{Warhaft,BlumBatch22}, chaotic mixing~\cite{Thiff08}, and turbulent aerosols~\cite{Mehlig}),
neuronal networks, weather science, and time-series analysis in general, while it may serve as a valuable complement for those interested in Lagrangian coherent structures~\cite{Haller,ChaotAdv17},
almost-invariant sets~\cite{Froy08}, or convective modes~\cite{Blachut,Froy21}.

As conveyed by the numerical tests on the Hamiltonian H\'enon map, the theory presented here may not necessarily be restricted to hyperbolic systems, but
rather extend to (noisy) mixed dynamics, when concerned with timescales that only involve the discrete part of the transport operator spectrum.    

On the other hand, the present treatment is limited to $i)$ systems whose densities and fields are Lebesgue integrable at all times, unlike for example noiseless strange attractors,
and $ii)$ featuring a real-valued second eigenvalue and eigenfunction for the transport operators. If, instead, the second eigenvalue is a complex conjugate pair, decay of correlations and convergence to the (conditionally)
invariant densities are not monotonic but oscillatory, and one can show that the field profile stemming from the
leading nontrivial terms in the expansions~(\ref{distrunc}) and~(\ref{adjdistrunc}) gets to  
depend on the choice of the observable, when the latter is real valued. 
This phenomenology is addressed in detail by a separate report~\cite{LippTherm}.

\appendix
\section{Appendix: Expansion truncations and timescales}
\label{AppTrunc}
In this section, I discuss the validity of the eigenfunction expansion in the
expressions~(\ref{distrunc})-(\ref{adjdistrunc}) for the distributions of both local and integrated observables, as well as its 
first-order truncation.

Suppose the model is an expanding or area-preserving map with no escape, like the Bernoulli or the cat map considered in section~\ref{numerics}. In that case, it always true that
\begin{equation}
\int_\pS \left[\Lop_t \rho\right](x) dx = \int_\pS \rho(x) dx
\,,
\end{equation} 
and likewise for the Koopman evolution $\Lop^\dagger_t$. As a consequence, setting 
$\rho(x)=\phi_j(x)$,  eigenfunction of the Perron-Frobenius (Koopman) operator, the previous 
becomes
\begin{equation}
e^{-\gamma_jt}\int_\pS \phi_j(x) dx = \int_\pS \phi_j(x) dx
\,,
\end{equation} 
which, in order to check out, needs either $\gamma_0=0$ (no escape), allowing
$\int_\pS \phi_0(x)dx \neq0$, or, for higher-order eigenfunctions, $\int_\pS \phi_j(x)dx =0$,
since $\gamma_j\neq0$ when $j>0$. This property allows the expansions~(\ref{distrunc})-(\ref{adjdistrunc}) to make sense when the observable $a(x)=C$ is  a constant, as they trivially 
reduce to the first term   
\begin{equation}
\hat{a}^{t}(x_0)  =
\frac{\sum_j e^{-\gamma_jt} \int_\pS a(x) \phi_j(x) \,dx}{\int_{\pS} \rho(x)}   =  
\frac{C \int_{\pS} \phi_0(x) \,dx}{\int_{\pS} \rho(x)}  = C
\,.
\end{equation}  
If we consider the effect of a squeezing map, instead, like in the adjoint evolution of the Bernoulli map, the present theory does not apply due to the non-smoothness of the eigenfunctions of the 
evolution operator. If the system has no phase space contraction but admits escape, then
the field $\hat{a}^t(x)$ is time dependent even when the initial value $a(x)=C$ is a constant.
That is because trajectories are escaping and the support of the field shrinks with time.  In that
case, the expansion
\begin{equation}
  \langle a^t \rangle(x_0)  = \frac{\sum_j^\infty \varphi_j(x_0) e^{-\gamma_j t} \int_\pS dx\, \phi_j(x)  a(x)} {\sum_j^\infty \varphi_j(x_0) e^{-\gamma_j t} \int_\pS dx\, \phi_j(x)}   
\end{equation}
does bear the nonzero terms $\int_\pS dx\, \phi_j(x)$ at the denominator, which are however expected to be small or $O(1)$ (normalization requires $\int_\pS dx\, \phi_0(x)=1$). This was numerically verified for the 
H\'enon map (not shown). For that reason, the approximations~(\ref{distrunc})-(\ref{adjdistrunc}) made for the
observables only feature $\varphi_0(x)$ ($\phi_0(x)$) at the denominator.         

Let us now shift our focus to the higher-order terms that would follow the truncations~(\ref{distrunc})  and~(\ref{adjdistrunc}). To the next order, the expansion is
\begin{equation}
\hat{a}^{t}(x_0) 
\simeq  \int_\pS a(x) \phi_0(x) \,dx   +  \frac{\varphi_1(x_0)}{\varphi_0(x_0)}e^{-(\gamma_1-\gamma_0)t} \int_\pS a(x) \phi_1(x) \,dx + 
 \frac{\varphi_2(x_0)}{\varphi_0(x_0)}e^{-(\gamma_2-\gamma_0)t} \int_\pS a(x) \phi_2(x) \,dx 
\,,
\label{distrunc2}
\end{equation}   
so that the relative magnitude of the second and third terms may be assessed by looking at their 
respective time-dependent prefactors
\begin{equation} 
c_i(t) = e^{-(\gamma_i-\gamma_0)t} \int_\pS a(x) \phi_i(x) \,dx 
\,.
\label{ci}
\end{equation}
\begin{figure}[tbh!]
\centerline{
\scalebox{.6}{\includegraphics{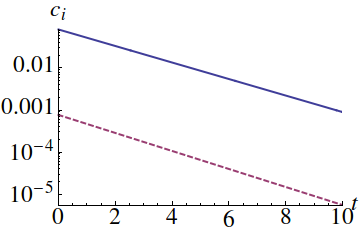}}}
\caption{(Log-scale) The prefactors $c_1(t)$ (solid line) and $c_2(t)$ (dashed line) as defined in equation~(\ref{ci}), computed for the perturbed cat map model and the observable $q^2$, versus time.} 
\label{CatInstFact}
\end{figure}
Figure~\ref{CatInstFact} shows the decay of both $c_1(t)$ and $c_2(t)$ in the simulations of the perturbed cat map and the observable $a(x)=q^2$, where the first term consistently exceeds the second by at least two orders of magnitudes over the time scale of relaxation of the system to statistical equilibrium, determined by the spectral gap (figures of the same order were found for the other observables considered in this work). It is noted that the second and third eigenfunctions $\varphi_1(x)$ and $\varphi_2(x)$ take values of the same order, both being $L^1-$normalized. Besides the fact that $\gamma_2>\gamma_1$, higher-order eigenfunctions tend to increasingly oscillate~\cite{BakerSpec,CatManSpec}, and thus the integral $\int_\pS a(x) \phi_i(x) dx$ decreases 
   in absolute value, as $i$ grows. As a consequence, the third eigenvalue $e^{-\gamma_2}$ of the transfer operator need not be much smaller than the second $e^{-\gamma_1}$ (in this calculation they only differ by $4\%$) in order for the truncation to the second term in the expansion to closely approximate the distribution $\hat{a}^t$ for much of the relaxation time to equilibrium or stationarity.    
 On the other hand, the first non-trivial contribution to the expansion~(\ref{distrunc2}), which in the cat map is found to be $O(10^{-1})$ at intermediate times for all the observables studied, is significant with respect to the leading term, that is of the order of unity. For the previous argument to hold for a wide class of models, it is necessary but not unreasonable to assume $a(x)$ as
smooth, relatively to the high-order eigenfunctions of the transfer operators. 

Next, let us examine integrated observables. Here the analysis is slightly more involved, since the factors 
$\int_\pS A^t(x) \phi_1(x) dx$ in the expansion 
\begin{equation}
 \hat{A}^t(x_0) 
\simeq  \int_\pS A^t(x) \phi_0(x) \,dx   +  \frac{\varphi_1(x_0)}{\varphi_0(x_0)}e^{-(\gamma_1-\gamma_0)t} \int_\pS A^t(x) \phi_1(x) \,dx\,, +
 \frac{\varphi_2(x_0)}{\varphi_0(x_0)}e^{-(\gamma_2-\gamma_0)t} \int_\pS A^t(x) \phi_2(x) \,dx\,, 
\label{Adistrunc2}
\end{equation}
are time dependent. If our observable $a(x)=g(x)$, that is a function of the coordinates (like diffusivity and kinetic energy, unlike the Lyapunov exponent), however, it is possible to go back to the definition~(\ref{intobs}) of integrated observable, rewrite it in terms of the Koopman operator, as  
\begin{equation}
\nonumber 
A^t(x)= \int_0^t a\left[f^\tau(x)\right]\,d\tau = \int_0^t [\Lop^\dagger_\tau a](x) d\tau
\,,
\end{equation}
and expand $[\Lop^\dagger_\tau a](x)$ in terms of the eigenspectrum of $\Lop^\dagger_\tau$. That way,
every factor multiplying the eigenfunction ratio $\varphi_i(x_0)/\varphi_0(x_0)$ in equation~(\ref{Adistrunc2}) is written as
\begin{equation}
 c_i(t) = e^{-(\gamma_i-\gamma_0)t} \sum_j \hat{b}_j
\frac{1-e^{-\gamma_j t}}{\gamma_j} \int_\pS   \varphi_j(x) \phi_i(x) \,dx =
  \hat{b}_i e^{-(\gamma_i-\gamma_0)t}
\frac{1-e^{-\gamma_i t}}{\gamma_i}
\,,
\label{IntegCi}
\end{equation}
for the bi-orthonormality of the eigenfunctions of $\Lop^\dagger$ and $\Lop$, with
$\hat{b}_i = \int_\pS \phi_i(x) a(x) dx$. It is immediately apparent the difference of the time-dependent
part of $c_i(t)$ here in comparison with equation~(\ref{ci}) for a local observable.
\begin{figure}[tbh!]
\centerline{
(a)\scalebox{.55}{\includegraphics{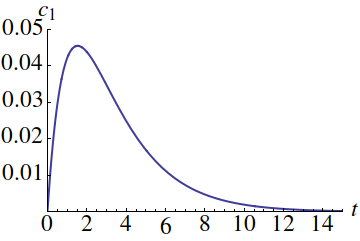}}
\hskip 0.1cm
(b)\scalebox{.55}{\includegraphics{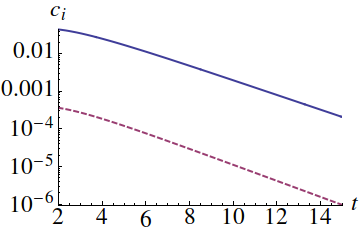}}
}
\caption{(a) The prefactor $c_1(t)$,  as defined in equation~(\ref{IntegCi}), computed for the perturbed cat map model and the integrated observable
$A^t(x) = \int_0^t d\tau q^2(f^\tau(x))$, versus time. (b) (Log-scale) The prefactors $c_1(t)$ (solid line) and $c_2(t)$ (dashed line) versus time, for the same system and integrated observable.} 
\label{CatIntegFact}
\end{figure}
Figure~\ref{CatIntegFact}(a) illustrates this behavior, and in particular it features a slower relaxation to
equilibrium for an integrated observable than for a local one. That explains why the  phase-space profiles of $A^t(x)$ still bear the signatures of the second eigenfunction for a relatively large value of $t$ in the snapshots of Figs.~\ref{CatInitt5},~\ref{CatFint5},~\ref{henobs},~\ref{henbobs}.
Like in the case of a local observable, I now compare the prefactors $c_1$ and $c_2$ of the second and third eigenfunctions respectively,  this time given by equation~(\ref{IntegCi}), in the example of the perturbed cat map, for the observable $A^t(x) = \int_0^t d\tau q^2(f^\tau(x))$. The outcome (Fig.~\ref{CatIntegFact}(b))
confirms the conclusions drawn for local observables, where the coefficient of the third term in the expansion is consistently at least two orders of magnitudes below that of the second at an intermediate time scale toward relaxation, and thus justifies the truncation that motivates the 
fields' observable-independent behavior claimed here. The contribution of the factor $\hat{b}_i = \int_\pS \phi_i(x) a(x) dx$ is again decisive, and so is the increasingly oscillatory behavior of $\phi_i(x)$ with the order $i$
of the eigenfunction.

\printcredits

\bibliographystyle{cas-model2-names}

\bibliography{PLA-refs}

\begin{thebibliography}{47}
\expandafter\ifx\csname natexlab\endcsname\relax\def\natexlab#1{#1}\fi
\providecommand{\url}[1]{\texttt{#1}}
\providecommand{\href}[2]{#2}
\providecommand{\path}[1]{#1}
\providecommand{\DOIprefix}{doi:}
\providecommand{\ArXivprefix}{arXiv:}
\providecommand{\URLprefix}{URL: }
\providecommand{\Pubmedprefix}{pmid:}
\providecommand{\doi}[1]{\href{http://dx.doi.org/#1}{\path{#1}}}
\providecommand{\Pubmed}[1]{\href{pmid:#1}{\path{#1}}}
\providecommand{\bibinfo}[2]{#2}
\ifx\xfnm\relax \def\xfnm[#1]{\unskip,\space#1}\fi
\bibitem[{Ahsan et~al.(2024)Ahsan, Dankowicz and Kuehn}]{Dankowicz}
\bibinfo{author}{Ahsan, Z.}, \bibinfo{author}{Dankowicz, H.},
  \bibinfo{author}{Kuehn, C.}, \bibinfo{year}{2024}.
\newblock \bibinfo{title}{Adjoint-based projections for uncertainty
  quantification near stochastically perturbed limit cycles and tori}.
\newblock \bibinfo{howpublished}{arXiv.2404.13429}.
\newblock \URLprefix \url{https://doi.org/10.48550/arXiv.2404.13429}.
\bibitem[{Alonso et~al.(2018)Alonso, M\'endez-Berm\'udez,
  Gonz\'alez-Mel\'endrez and Moreno}]{Moreno}
\bibinfo{author}{Alonso, L.}, \bibinfo{author}{M\'endez-Berm\'udez, J.A.},
  \bibinfo{author}{Gonz\'alez-Mel\'endrez, A.}, \bibinfo{author}{Moreno, Y.},
  \bibinfo{year}{2018}.
\newblock \bibinfo{title}{Weighted random-geometric and random-rectangular
  graphs: spectral and eigenfunction properties of the adjacency matrix}.
\newblock \bibinfo{journal}{Journal of Complex Networks} \bibinfo{volume}{6},
  \bibinfo{pages}{753}.
\bibitem[{Altmann et~al.(2013)Altmann, Portela and T\'el}]{AltLeak}
\bibinfo{author}{Altmann, E.G.}, \bibinfo{author}{Portela, J.S.E.},
  \bibinfo{author}{T\'el, T.}, \bibinfo{year}{2013}.
\newblock \bibinfo{title}{Leaking chaotic systems}.
\newblock \bibinfo{journal}{Reviews of Modern Physics} \bibinfo{volume}{85},
  \bibinfo{pages}{869}.
\bibitem[{Aref et~al.(2017)}]{ChaotAdv17}
\bibinfo{author}{Aref, H.}, et~al., \bibinfo{year}{2017}.
\newblock \bibinfo{title}{Frontiers of chaotic advection}.
\newblock \bibinfo{journal}{Reviews of Modern Physics} \bibinfo{volume}{89},
  \bibinfo{pages}{025007}.
\bibitem[{Arnold and Avez(1968)}]{Arnold}
\bibinfo{author}{Arnold, V.I.}, \bibinfo{author}{Avez, A.},
  \bibinfo{year}{1968}.
\newblock \bibinfo{title}{Ergodic Problems of Classical Mechanics}.
\newblock \bibinfo{publisher}{Benjiamin}, \bibinfo{address}{New York}.
\bibitem[{Bec et~al.(2024)Bec, Gustavsson and Mehlig}]{Mehlig}
\bibinfo{author}{Bec, J.}, \bibinfo{author}{Gustavsson, K.},
  \bibinfo{author}{Mehlig, B.}, \bibinfo{year}{2024}.
\newblock \bibinfo{title}{Statistical models for the dynamics of heavy
  particles in turbulence}.
\newblock \bibinfo{journal}{Annual Review of Fluid Mechanics}
  \bibinfo{volume}{56}, \bibinfo{pages}{189}.
\bibitem[{Bedrossian et~al.(2022)Bedrossian, Blumenthal and
  Punshon-Smith}]{BlumBatch22}
\bibinfo{author}{Bedrossian, J.}, \bibinfo{author}{Blumenthal, A.},
  \bibinfo{author}{Punshon-Smith, S.}, \bibinfo{year}{2022}.
\newblock \bibinfo{title}{The {B}atchelor spectrum of passive scalar turbulence
  in stochastic fluid mechanics at fixed {R}eynolds number}.
\newblock \bibinfo{journal}{Communications on Pure and Applied Mathematics}
  \bibinfo{volume}{75}, \bibinfo{pages}{1237}.
\bibitem[{Blachut and Balasuriya(2024)}]{Blachut}
\bibinfo{author}{Blachut, C.}, \bibinfo{author}{Balasuriya, S.},
  \bibinfo{year}{2024}.
\newblock \bibinfo{title}{Convective modes reveal the incoherence of the
  {S}outhern {P}olar {V}ortex}.
\newblock \bibinfo{journal}{Scientific Reports} \bibinfo{volume}{14},
  \bibinfo{pages}{966}.
\bibitem[{Blank et~al.(2002)Blank, Keller and Liverani}]{BKL02}
\bibinfo{author}{Blank, M.}, \bibinfo{author}{Keller, G.},
  \bibinfo{author}{Liverani, C.}, \bibinfo{year}{2002}.
\newblock \bibinfo{title}{Ruelle-{P}erron-{F}robenius spectrum for {A}nosov
  maps}.
\newblock \bibinfo{journal}{Nonlinearity} \bibinfo{volume}{15},
  \bibinfo{pages}{1905}.
\bibitem[{Bollt and Ross(2021)}]{RossBollt}
\bibinfo{author}{Bollt, E.M.}, \bibinfo{author}{Ross, S.D.},
  \bibinfo{year}{2021}.
\newblock \bibinfo{title}{Is the finite-time {L}yapunov exponent field a
  {K}oopman eigenfunction?}
\newblock \bibinfo{journal}{Mathematics} \bibinfo{volume}{9},
  \bibinfo{pages}{2731}.
\bibitem[{Bratteli and Jorgensen(2002)}]{BratJorg}
\bibinfo{author}{Bratteli, O.}, \bibinfo{author}{Jorgensen, P.},
  \bibinfo{year}{2002}.
\newblock \bibinfo{title}{The Transfer Operator and Perron-Frobenius Theory}.
\newblock \bibinfo{publisher}{Birk\"auser}, \bibinfo{address}{Basel}.
\bibitem[{Budi\u{s}i\'c et~al.(2012)Budi\u{s}i\'c, Mohr and
  Mezi\'c}]{Koopmanism}
\bibinfo{author}{Budi\u{s}i\'c, M.}, \bibinfo{author}{Mohr, R.},
  \bibinfo{author}{Mezi\'c, I.}, \bibinfo{year}{2012}.
\newblock \bibinfo{title}{Applied {K}oopmanism}.
\newblock \bibinfo{journal}{Chaos} \bibinfo{volume}{22},
  \bibinfo{pages}{047510}.
\bibitem[{Chappell and Tanner(2013)}]{ChapTan}
\bibinfo{author}{Chappell, D.}, \bibinfo{author}{Tanner, G.},
  \bibinfo{year}{2013}.
\newblock \bibinfo{title}{Solving the {L}iouville equation via a boundary
  element method}.
\newblock \bibinfo{journal}{Journal of Computational Physics}
  \bibinfo{volume}{234}, \bibinfo{pages}{487}.
\bibitem[{Cvitanovi\'c et~al.(2020)Cvitanovi\'c, Artuso, Mainieri, Tanner and
  Vattay}]{dasBuch}
\bibinfo{author}{Cvitanovi\'c, P.}, \bibinfo{author}{Artuso, R.},
  \bibinfo{author}{Mainieri, R.}, \bibinfo{author}{Tanner, G.},
  \bibinfo{author}{Vattay, G.}, \bibinfo{year}{2020}.
\newblock \bibinfo{title}{Chaos: Classical and Quantum}.
\newblock \bibinfo{publisher}{Niels Bohr Institute},
  \bibinfo{address}{Copenhagen}.
\bibitem[{Cvitanovi\'c and Lippolis(2012)}]{CviLip12}
\bibinfo{author}{Cvitanovi\'c, P.}, \bibinfo{author}{Lippolis, D.},
  \bibinfo{year}{2012}.
\newblock \bibinfo{title}{Knowing when to stop: how noise frees us from
  determinism}.
\newblock \bibinfo{journal}{AIP Conference Proceedings} \bibinfo{volume}{1468},
  \bibinfo{pages}{82}.
\bibitem[{Dellnitz and Junge(1999)}]{DellJunge}
\bibinfo{author}{Dellnitz, M.}, \bibinfo{author}{Junge, O.},
  \bibinfo{year}{1999}.
\newblock \bibinfo{title}{On the approximation of complicated dynamical
  behavior}.
\newblock \bibinfo{journal}{SIAM Journal of Numerical Analysis}
  \bibinfo{volume}{36}, \bibinfo{pages}{491}.
\bibitem[{Driebe(1999)}]{Driebe}
\bibinfo{author}{Driebe, D.}, \bibinfo{year}{1999}.
\newblock \bibinfo{title}{Fully Chaotic Maps and Broken Time Symmetry}.
\newblock \bibinfo{publisher}{Springer}, \bibinfo{address}{Dodrecht}.
\bibitem[{D'Souza et~al.(2023)D'Souza, di~Bernardo and Liu}]{Raissa}
\bibinfo{author}{D'Souza, R.M.}, \bibinfo{author}{di~Bernardo, M.},
  \bibinfo{author}{Liu, Y.Y.}, \bibinfo{year}{2023}.
\newblock \bibinfo{title}{Controlling complex networks with complex nodes}.
\newblock \bibinfo{journal}{Nature Reviews Physics} \bibinfo{volume}{5},
  \bibinfo{pages}{250}.
\bibitem[{Ermann and Shepelyansky(2012)}]{ErmShep}
\bibinfo{author}{Ermann, L.}, \bibinfo{author}{Shepelyansky, D.L.},
  \bibinfo{year}{2012}.
\newblock \bibinfo{title}{The {A}rnold cat map, the {U}lam method, and time
  reversal}.
\newblock \bibinfo{journal}{Physica D} \bibinfo{volume}{241},
  \bibinfo{pages}{514}.
\bibitem[{Faure and Roy(2006)}]{FaureRoy}
\bibinfo{author}{Faure, F.}, \bibinfo{author}{Roy, N.}, \bibinfo{year}{2006}.
\newblock \bibinfo{title}{Ruelle–{P}ollicott resonances for real analytic
  hyperbolic maps}.
\newblock \bibinfo{journal}{Nonlinearity} \bibinfo{volume}{19},
  \bibinfo{pages}{1233}.
\bibitem[{Fox(1997)}]{Fox}
\bibinfo{author}{Fox, R.}, \bibinfo{year}{1997}.
\newblock \bibinfo{title}{Construction of the {J}ordan basis for the {B}aker
  map}.
\newblock \bibinfo{journal}{Chaos} \bibinfo{volume}{7}, \bibinfo{pages}{254}.
\bibitem[{Froyland(2007)}]{Froy07}
\bibinfo{author}{Froyland, G.}, \bibinfo{year}{2007}.
\newblock \bibinfo{title}{On {U}lam approximation of the isolated spectrum and
  eigenfunctions of hyperbolic maps}.
\newblock \bibinfo{journal}{Discrete \& Continuous Dynamical Systems}
  \bibinfo{volume}{17}, \bibinfo{pages}{671}.
\bibitem[{Froyland(2008)}]{Froy08}
\bibinfo{author}{Froyland, G.}, \bibinfo{year}{2008}.
\newblock \bibinfo{title}{Unwrapping eigenfunctions to discover the geometry of
  almost-invariant sets in hyperbolic maps}.
\newblock \bibinfo{journal}{Physica D} \bibinfo{volume}{237},
  \bibinfo{pages}{840}.
\bibitem[{Froyland(2013)}]{Froy13}
\bibinfo{author}{Froyland, G.}, \bibinfo{year}{2013}.
\newblock \bibinfo{title}{An analytic framework for identifying finite-time
  coherent sets in time-dependent dynamical systems}.
\newblock \bibinfo{journal}{Physica D} , \bibinfo{pages}{1}.
\bibitem[{Froyland et~al.(2021)Froyland, Giannakis, Lintner, Pike and
  Slawinska}]{Froy21}
\bibinfo{author}{Froyland, G.}, \bibinfo{author}{Giannakis, D.},
  \bibinfo{author}{Lintner, B.R.}, \bibinfo{author}{Pike, M.},
  \bibinfo{author}{Slawinska, J.}, \bibinfo{year}{2021}.
\newblock \bibinfo{title}{Spectral analysis of climate dynamics with
  operator-theoretic approaches}.
\newblock \bibinfo{journal}{Nature Communications} \bibinfo{volume}{12},
  \bibinfo{pages}{6570}.
\bibitem[{Gaspard(1999)}]{Gaspard}
\bibinfo{author}{Gaspard, P.}, \bibinfo{year}{1999}.
\newblock \bibinfo{title}{Chaos, Scattering, and Statistical Mechanics}.
\newblock \bibinfo{publisher}{Cambridge University Press},
  \bibinfo{address}{Cambridge}.
\bibitem[{Haller(2015)}]{Haller}
\bibinfo{author}{Haller, G.}, \bibinfo{year}{2015}.
\newblock \bibinfo{title}{Lagrangian coherent structures}.
\newblock \bibinfo{journal}{Annual Review of Fluid Mechanics}
  \bibinfo{volume}{47}, \bibinfo{pages}{137}.
\bibitem[{Hasegawa and Driebe(1994)}]{BakerSpec}
\bibinfo{author}{Hasegawa, H.H.}, \bibinfo{author}{Driebe, D.J.},
  \bibinfo{year}{1994}.
\newblock \bibinfo{title}{Intrinsic irreversibility and the validity of the
  kinetic description of chaotic systems}.
\newblock \bibinfo{journal}{Physical Review E} \bibinfo{volume}{50},
  \bibinfo{pages}{1781}.
\bibitem[{Hasegawa et~al.(2003)Hasegawa, Driebe and Li}]{CatManSpec}
\bibinfo{author}{Hasegawa, H.H.}, \bibinfo{author}{Driebe, D.J.},
  \bibinfo{author}{Li, C.B.}, \bibinfo{year}{2003}.
\newblock \bibinfo{title}{Spectral decomposition of the stretching dynamics of
  the {A}rnold cat map}.
\newblock \bibinfo{journal}{Physics Letters A} \bibinfo{volume}{319},
  \bibinfo{pages}{290}.
\bibitem[{Kapfer and Krauth(2017)}]{Krauth}
\bibinfo{author}{Kapfer, S.C.}, \bibinfo{author}{Krauth, W.},
  \bibinfo{year}{2017}.
\newblock \bibinfo{title}{Irreversible local {M}arkov chains with rapid
  convergence towards equilibrium}.
\newblock \bibinfo{journal}{Physical Review Letters} \bibinfo{volume}{119},
  \bibinfo{pages}{240603}.
\bibitem[{Klus et~al.(2020)Klus, N\"uske, Peitz, Niemann, Clementi and
  Sch\"utte}]{Klus}
\bibinfo{author}{Klus, S.}, \bibinfo{author}{N\"uske, F.},
  \bibinfo{author}{Peitz, S.}, \bibinfo{author}{Niemann, J.H.},
  \bibinfo{author}{Clementi, C.}, \bibinfo{author}{Sch\"utte, C.},
  \bibinfo{year}{2020}.
\newblock \bibinfo{title}{Data-driven approximation of the {K}oopman generator:
  Model reduction, system identification, and control}.
\newblock \bibinfo{journal}{Physica D} \bibinfo{volume}{406},
  \bibinfo{pages}{132416}.
\bibitem[{Koopman(1931)}]{Koopman}
\bibinfo{author}{Koopman, B.O.}, \bibinfo{year}{1931}.
\newblock \bibinfo{title}{Hamiltonian systems and transformations in {H}ilbert
  spaces}.
\newblock \bibinfo{journal}{Proceedings of the National Academy of Sciences}
  \bibinfo{volume}{17}, \bibinfo{pages}{315}.
\bibitem[{Kringelbach and Deco(2020)}]{Deco}
\bibinfo{author}{Kringelbach, M.L.}, \bibinfo{author}{Deco, G.},
  \bibinfo{year}{2020}.
\newblock \bibinfo{title}{Brain states and transitions: Insights from
  computational neuroscience}.
\newblock \bibinfo{journal}{Cell Reports} \bibinfo{volume}{32},
  \bibinfo{pages}{108128}.
\bibitem[{Lasota and MacKey(1994)}]{LasMac}
\bibinfo{author}{Lasota, A.}, \bibinfo{author}{MacKey, M.},
  \bibinfo{year}{1994}.
\newblock \bibinfo{title}{Chaos, Fractals, and Noise: Stochastic Aspects of
  Dynamics}.
\newblock \bibinfo{publisher}{Springer}, \bibinfo{address}{New York}.
\bibitem[{Lippolis(2024)}]{LippTherm}
\bibinfo{author}{Lippolis, D.}, \bibinfo{year}{2024}.
\newblock \bibinfo{title}{Thermodynamics of chaotic relaxation processes}.
\newblock \bibinfo{journal}{Physical Review E} \bibinfo{volume}{110},
  \bibinfo{pages}{024215}.
\bibitem[{Lippolis et~al.(2021)Lippolis, Shudo, Yoshida and
  Yoshino}]{ClassScars}
\bibinfo{author}{Lippolis, D.}, \bibinfo{author}{Shudo, A.},
  \bibinfo{author}{Yoshida, K.}, \bibinfo{author}{Yoshino, H.},
  \bibinfo{year}{2021}.
\newblock \bibinfo{title}{Scarring in classical chaotic dynamics with noise}.
\newblock \bibinfo{journal}{Physical Review E} \bibinfo{volume}{103},
  \bibinfo{pages}{L050202}.
\bibitem[{Liu and Haller(2004)}]{LiuHaller}
\bibinfo{author}{Liu, W.}, \bibinfo{author}{Haller, G.}, \bibinfo{year}{2004}.
\newblock \bibinfo{title}{Strange eigenmodes and decay of variance in the
  mixing diffusive tracers}.
\newblock \bibinfo{journal}{Physica D} \bibinfo{volume}{188},
  \bibinfo{pages}{1}.
\bibitem[{Maiocchi et~al.(2022)Maiocchi, Lucarini and Gritsun}]{Lucar}
\bibinfo{author}{Maiocchi, C.C.}, \bibinfo{author}{Lucarini, V.},
  \bibinfo{author}{Gritsun, A.}, \bibinfo{year}{2022}.
\newblock \bibinfo{title}{Decomposing the dynamics of the {L}orenz 1963 model
  using unstable periodic orbits: Averages, transitions, and quasi-invariant
  sets}.
\newblock \bibinfo{journal}{Chaos} \bibinfo{volume}{32},
  \bibinfo{pages}{033129}.
\bibitem[{Pierrehumbert(1994)}]{Pierrehumb}
\bibinfo{author}{Pierrehumbert, R.T.}, \bibinfo{year}{1994}.
\newblock \bibinfo{title}{Tracer microstructure in the large-eddy dominated
  regime}.
\newblock \bibinfo{journal}{Chaos, Solitons, and Fractals} \bibinfo{volume}{4},
  \bibinfo{pages}{1091}.
\bibitem[{Pikovsky and Popovych(2003)}]{PikPop}
\bibinfo{author}{Pikovsky, A.}, \bibinfo{author}{Popovych, O.},
  \bibinfo{year}{2003}.
\newblock \bibinfo{title}{Persistent patterns in deterministic mixing flows}.
\newblock \bibinfo{journal}{Europhysics Letters} \bibinfo{volume}{61},
  \bibinfo{pages}{625}.
\bibitem[{Risken(1996)}]{Risken}
\bibinfo{author}{Risken, H.}, \bibinfo{year}{1996}.
\newblock \bibinfo{title}{The Fokker-Planck Equation}.
\newblock \bibinfo{publisher}{Springer}, \bibinfo{address}{Berlin}.
\bibitem[{Slipantschuk et~al.(2013)Slipantschuk, Bandtlow and Just}]{Slipan13}
\bibinfo{author}{Slipantschuk, J.}, \bibinfo{author}{Bandtlow, O.F.},
  \bibinfo{author}{Just, W.}, \bibinfo{year}{2013}.
\newblock \bibinfo{title}{On the relation between {L}yapunov exponents and
  exponential decay of correlations}.
\newblock \bibinfo{journal}{Journal of Physics A: Mathematical and Theoretical}
  \bibinfo{volume}{46}, \bibinfo{pages}{075101}.
\bibitem[{Souza(2023)}]{Andre}
\bibinfo{author}{Souza, A.N.}, \bibinfo{year}{2023}.
\newblock \bibinfo{title}{Transforming butterflies into graphs: Statistics of
  chaotic and turbulent systems}.
\newblock \bibinfo{howpublished}{arXiv.2304.03362}.
\newblock \URLprefix \url{https://doi.org/10.48550/arXiv.2304.03362}.
\bibitem[{Thiffeault(2007)}]{Thiff08}
\bibinfo{author}{Thiffeault, J.L.}, \bibinfo{year}{2007}.
\newblock \bibinfo{title}{Scalar decay in chaotic mixing}.
\newblock \bibinfo{journal}{Lecture Notes in Physics} \bibinfo{volume}{744},
  \bibinfo{pages}{3}.
\bibitem[{Ulam(1960)}]{Ulam}
\bibinfo{author}{Ulam, S.M.}, \bibinfo{year}{1960}.
\newblock \bibinfo{title}{A Collection of Mathematical Problems}.
\newblock \bibinfo{publisher}{Interscience}, \bibinfo{address}{New york}.
\bibitem[{Warhaft(2000)}]{Warhaft}
\bibinfo{author}{Warhaft, Z.}, \bibinfo{year}{2000}.
\newblock \bibinfo{title}{Passive scalars in turbulent flows}.
\newblock \bibinfo{journal}{Annual Review of Fluid Mechanics}
  \bibinfo{volume}{32}, \bibinfo{pages}{203}.
\bibitem[{Yoshida et~al.(2021)Yoshida, Yoshino, Shudo and Lippolis}]{Kensuke}
\bibinfo{author}{Yoshida, K.}, \bibinfo{author}{Yoshino, H.},
  \bibinfo{author}{Shudo, A.}, \bibinfo{author}{Lippolis, D.},
  \bibinfo{year}{2021}.
\newblock \bibinfo{title}{Eigenfunctions of the {P}erron--{F}robenius operator
  and the finite-time {L}yapunov exponents in uniformly hyperbolic
  area-preserving maps}.
\newblock \bibinfo{journal}{Journal of Physics A: Mathematical and Theoretical}
  \bibinfo{volume}{54}, \bibinfo{pages}{285701}.

\end{thebibliography}

\end{document}